\documentclass[MTech]{iitmdiss}

\usepackage{graphicx}
\usepackage{psfrag}

\usepackage{subcaption}
\usepackage{stfloats}
\usepackage{epsfig}

\usepackage{amsmath} 
\usepackage{amsfonts}

\usepackage[authoryear]{natbib}
\usepackage{hyperref} 
\setcitestyle{square}

\usepackage{tocbibind}

\usepackage{algorithm}
\usepackage[noend]{algpseudocode}

\usepackage{tikz}
\usetikzlibrary{shapes,arrows}
\usetikzlibrary{calc, positioning}

\tikzset{
  block/.style    = {draw, rectangle, minimum height = 3em, minimum width = 5em},
  smallblock/.style    = {draw, rectangle, minimum height = 3em, minimum width = 4em},
  longblock/.style    = {draw, rectangle, minimum height = 7em, minimum width = 1em},
  sum/.style      = {draw, circle, node distance = 1cm},
  input/.style    = {coordinate},
  output/.style   = {coordinate}
}
\newcommand{\sampler}[3]{
    \node[right=#1 of #2](#3){};
    \draw[-] ($(#3.south west)+(0mm,1.1mm)$) -- ($(#3.north east)+(0,1mm)$);
}

\DeclareMathOperator*{\argmax}{arg\,max}

\begin{document}
\setcitestyle{authoryear, open={[}, close={]}}
\nocite{*}
\newcommand{\titleallcaps}{FASTER-THAN-NYQUIST SIGNALLING - THEORETICAL LIMITS ON CAPACITY AND TECHNIQUES TO APPROACH CAPACITY} 
\newcommand{\titlecamelcaps}{Faster-Than-Nyquist Signalling - Theoretical Limits on Capacity and Techniques to Approach Capacity} 
\setcounter{secnumdepth}{4}


\title{\titleallcaps}

\author{Sathwik Chadaga P}

\date{June, 2020}
\department{ELECTRICAL ENGINEERING}

\maketitle

\certificate

\vspace*{0.5in}

\noindent This is to certify that the thesis titled {\bf \titlecamelcaps}, submitted by {\bf Sathwik Chadaga P}, to the Indian Institute of Technology Madras, for
the award of the degree of {\bf Bachelor and Master of Technology}, is a bona fide record of the research work done by him under our supervision.  The contents of this thesis, in full or in parts, have not been submitted to any other Institute or University for the award of any degree or diploma.

\vspace*{1.5in}

\begin{singlespacing}
\hspace*{-0.25in}
\parbox{3in}{
\noindent {\bf Dr. David Koilpillai} \\
\noindent Research Guide \\ 
\noindent Professor \\
\noindent Department of Electrical Engineering\\
\noindent IIT Madras, 600036 \\
\noindent Place: Chennai\\
Date: June 19, 2020 
} 
\end{singlespacing}

\chapter*{\centerline{ACKNOWLEDGEMENTS}}
\pagenumbering{roman}
\setcounter{page}{1} 
\addcontentsline{toc}{chapter}{ACKNOWLEDGEMENTS}

I am grateful to my thesis advisor Prof. David Koilpillai from the Department of Electrical Engineering at IIT Madras for giving me the opportunity to take up this project under his supervision. I wish to thank him for
his support and guidance regarding my thesis. His exceptional teaching of courses like Wireless Communication and Multirate Digital Signal Processing has helped me build a strong foundation in the field of communications and signal processing. I also thank him for assisting my research career growth by encouraging me to take part in unique programs like conferences, research internships, and teaching experiences. 

I also thank Prof. Nambi Seshadri from the ECE Department at the University of California San Diego for advising me on this project. His insights and suggestions have played a significant part in the progress of my project. Periodic meetings with him have ensured that I have been able to clarify doubts immediately and continue with my work. Not only has he improved my creative and critical thinking abilities related to my research work, he has also encouraged me to look at many interesting real world problems from fresh and unique perspectives. I have learned a lot from him as a researcher as well as a professional.

I thank Akansh Jain, class of 2019, for encouraging me to take part in this project initially. The foundations of the research work in this thesis were laid by him. His robust skills in mathematics and thorough understanding of communication systems has helped me strengthen my knowledge. I am grateful for his constant support and help towards my research even after his graduation.

I also thank my lab mates Shishir, Romil, Narendra, Major Arun Singh Pundir, and Major Sanjay Haojam for creating a cheerful and scholarly research environment in our lab. I will cherish the discussion sessions we used to have over tea and the puzzle solving sessions on the lab whiteboard.
 
Lastly, but definitely not the least, I would like to thank my parents for consistently supporting and guiding me through my life and studies.

\chapter*{\centerline{ABSTRACT}}
\addcontentsline{toc}{chapter}{ABSTRACT}

\noindent KEYWORDS: \hspace*{0.5em} \parbox[t]{4.4in}{Faster-Than-Nyquist; Nyquist Theorem; Nyquist Criterion; Inter Symbol Interference; Pulse Shaping; Capacity; Throughput; Power Allocation; Water-Filling; Adaptive Loading.}

\vspace*{24pt}

\noindent Faster-Than-Nyquist (FTN) Signalling is a non-orthogonal transmission scheme that violates the Nyquist zero-ISI criterion providing higher throughput and better spectral efficiency than a Nyquist transmission scheme.
In this thesis, the inter symbol interference (ISI) introduced by FTN signalling is studied, and conditions on pulse shapes and $\tau$ (time acceleration factor) are derived so that the ISI can be avoided completely. 
Further, these conditions are reinforced by investigating the theoretical limits on the capacities of FTN systems. 
Finally, the use of power allocation and adaptive loading techniques are explored in reducing the effect of ISI  and increasing the throughput of orthogonal frequency division multiplexing (OFDM) FTN systems. The implementation of these techniques and simulation results are also demonstrated.


\begin{singlespace}
\tableofcontents


\listoffigures
\end{singlespace}


\chapter*{\centerline{ABBREVIATIONS}}
\addcontentsline{toc}{chapter}{ABBREVIATIONS}

\noindent 
\begin{tabbing}
xxxxxxxxxxx \= xxxxxxxxxxxxxxxxxxxxxxxxxxxxxxxxxxxxxxxxxxxxxxxx \kill

\textbf{AWGN} \> Additive White Gaussian Noise \\

\textbf{BER} \> Bit Error Rate \\
\textbf{BPSK} \> Binary Phase Shift Keying \\

\textbf{CTFT} \> Continuous Time Fourier Transform \\
\textbf{CP} \> Cyclic Prefix \\

\textbf{DFT} \> Discrete Fourier Transform \\
\textbf{DTFT} \> Discrete Time Fourier Transform \\

\textbf{FFT} \> Fast Fourier Transform \\
\textbf{FTN}   \> Faster Than Nyquist \\

\textbf{GTMH} \> G To Minus Half \\

\textbf{IDFT} \> Inverse Discrete Fourier Transform \\
\textbf{IFFT} \> Inverse Fast Fourier Transform \\
\textbf{i.i.d.} \> Independent and Identically Distributed \\
\textbf{ISI} \> Inter Symbol Interference \\

\textbf{LDPC} \> Low Density Parity Check \\

\textbf{MLSE} \> Maximum Likelihood Sequence Estimation \\
\textbf{MMSE} \> Minimum Mean Square Error \\

\textbf{OFDM} \>  Orthogonal Frequency Division Multiplexing\\

\textbf{PSD} \> Power Spectral Density \\

\textbf{QAM} \> Quadrature Amplitude Modulation \\
\textbf{QPSK} \> Quadrature Phase Shift Keying \\

\textbf{RC}  \> Raised Cosine \\
\textbf{rect} \> Rectangle \\

\textbf{SNR} \> Signal to Noise Ratio \\
\textbf{SRRC} \> Square Root Raised Cosine \\
\textbf{SVD} \> Singular Value Decomposition \\

\textbf{tri} \> Triangle \\

\textbf{w.r.t.} \> With Respect To

\end{tabbing}


\chapter*{\centerline{NOTATION}}
\addcontentsline{toc}{chapter}{NOTATION}

\noindent 
\begin{singlespace}
\begin{tabbing}
xxxxxxxxxxx \= xxxxxxxxxxxxxxxxxxxxxxxxxxxxxxxxxxxxxxxxxxxxxxxx \kill

$b[n]$  \> Transmit bits \\
$s[n]$ \> Transmit symbols \\
$r(t)$ \> Received waveform \\

$\eta$ \> White noise \\
$\eta'$ \> Coloured noise \\

$W$ \> Bandwidth \\
$T$ \> Nyquist rate  \\
$\tau$ \> FTN time acceleration factor \\

$h_{TX}(t)$ \> Transmit modulating pulse \\
$h_{RX}(t)$ \> Receiver matched filter response \\
$h_{SRRC}(t)$ \> SRRC pulse \\
$h_{RC}(t)$ \> RC pulse \\
$\alpha$ \> Roll-off of SRRC pulse \\

$\mathcal{H}(f)$ \> CTFT of a signal $h(t)$ \\
$H(e^{j\omega})$ \> DTFT of a signal $h(t)$ \\
$h(t)|_T$ \> $h(t)$ sampled at $T$ \\
$h_T[n]$ \> $h(t)$ sampled at rate $T$ \\
$h[n]$ \> $h(t)$ sampled at rate $\tau T$ unless otherwise specified \\

$C_{flat}$ \> Capacity of continuous time system whose waveform has square PSD \\
$C_{non-flat}$ \> Capacity of continuous time system whose waveform has non-square PSD \\
$C_{DT}$ \> Capacity of discrete time system \\
$C_{FTN}$ \> Capacity of FTN system \\
$C_{FTN,SRRC}$ \> Capacity of FTN system when the modulating pulse is SRRC \\
$C_{FTN,rect}$ \>  Capacity of FTN system when the modulating pulse is rectangular \\

\end{tabbing}
\end{singlespace}


\chapter{INTRODUCTION}
\label{chap:introduction}

\begin{figure}[h]
\hspace{-1cm}
{\scriptsize
\begin{tikzpicture}[auto, >=latex']
\draw
    node [input] (input1) {} 
    node [block, right of=input1, node distance=1.75cm] (mod) {Modulation}
    node [block, right of=mod, node distance=3.25cm] (tx_pulse) {$h_{TX}(t)$}
    node [sum, right of=tx_pulse, node distance=1.75cm] (add_noise) {+}
    	node at (6.75, 1) [input] (input2) {}
    node [block, right of=add_noise, node distance=2cm] (matched_filter) {$h_{RX}(t)$}
    node [input, right of=matched_filter] (sampler) {}
    node [block, right of=sampler, node distance=2.5cm] (white) {Whitening}
    node [block, right of=white, node distance=2.5cm] (demod) {Demod}
	node [output, right of=demod, node distance=1.75cm] (output1){}
	;
	
	\sampler{0.75cm}{matched_filter}{sampler}
	\node at (10.5,-0.25) {rate $T$};
	
	\draw[->](input1) -- node {$b[n]$}(mod);
	\draw[->] (mod) -- node {$s[n]$} node [below] {spaced $T$} (tx_pulse);
	\draw[->](tx_pulse) -- node {}(add_noise);
	\draw[->](add_noise) -- node {$r(t)$}(matched_filter);
		\draw[->](input2) -- node {$\eta(t)$}(add_noise);
	\draw[-] (matched_filter) -- node {$y(t)$}(sampler);
	\draw[->](sampler) -- node {$y[n]$}(white);
 	\draw[->](white) -- node {$z[n]$} (demod);
	\draw[->](demod) -- node {$\hat{b}[n]$} (output1);
\end{tikzpicture}}
\caption{Block diagram of a modern digital communication system.}
\label{fig:general_bd_intro}
\end{figure}
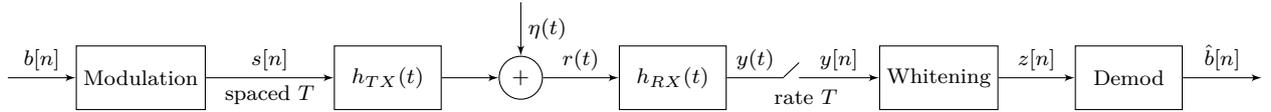

\pagenumbering{arabic}
The block diagram of a modern digital communication system is given in Fig. \ref{fig:general_bd_intro}.
In this digital communication system, the information bits $b[n]$ to be transmitted are mapped on to symbols $s[n]$ in the (I,Q) space (like M-level QAM) and are passed at a rate $T$ through a pulse shaping filter $h_{TX}(t)$ to obtain a continuous-time waveform. This waveform is up-converted and transmitted through the transmitter antenna and through an AWGN channel. The received waveform at the receiver antenna is down-converted to get the complex baseband received signal $r(t)$. The baseband received signal is passed through a filter $h_{RX}(t)$ that is matched to the transmit pulse $h_{TX}(t)$. The output of the matched filter is sampled at rate $T$ to obtain the received symbols, using which the transmitted information bits are estimated.

\section{Pulse Shaping in Linear Modulation}
\label{sec:pulse_shaping}
The pulse shaping $h_{TX}(t)$ is a linear modulation process and is of fundamental importance for communication over bandlimited channels, as it ensures that the transmit signal is bandlimited. The complex baseband received waveform for linear modulation can be written as
\begin{equation}
r(t) = \sum_m s[m]h_{TX}(t-mT) + \eta(t)
\end{equation}
where, $\eta[n]$ is the noise added by the AWGN channel. Note that throughout this thesis, an ideal single-tap channel is assumed. This signal is passed through a filter that is matched to the transmit pulse. The output of the matched filter is given by
\begin{equation}
y(t) = r(t)*h_{RX}(t) =  \sum_m s[m](h_{TX}*h_{RX})(t-m T) + \eta'(t)
\end{equation}
where, $*$ represents convolution and $\eta'(t)$ is the noise coloured due to convolution with the matched filter. This is passed through the sampler at rate $ T$ and the output of this  sampler is
\begin{equation}
y[n] = y(n T) = \sum_m s[m](h_{TX}*h_{RX})(n T-m T) + \eta'(n T).
\end{equation}
Simplifying,
\begin{equation}
\begin{split}
y[n] &= \sum_m s[m](h_{TX}*h_{RX})(nT-mT) + \eta'(nT)\\
&= \sum_m s[n-m](h_{TX}*h_{RX})(mT) + \eta'(nT)\\
&= s[n]*(h_{TX}*h_{RX})(nT) + \eta'(nT)
\end{split}
\label{eq:final_system_op}
\end{equation}
where, $(h_{TX}*h_{RX})(nT)$ is the sampled sequence $(h_{TX}*h_{RX})(t)|_{T}$.

The pulse shaping ensures that the transmit signal is bandlimited. But in the process, the overall process of pulse shaping and matched filtering can potentially introduce introduce ISI between neighbouring symbols. This is evident from (\ref{eq:final_system_op}), the overall response from $s[n]$ to $y[n]$ is given by $(h_{TX}*h_{RX})(nT)$. If this response is not Dirac delta, then there is ISI between neighbouring symbols. This leads to a criterion on the pulse shape that ensures that there is no ISI in the system. This criterion is called the Nyquist zero-ISI criterion. 

\section{The Nyquist Zero-ISI Criterion}
\label{sec:nyquist_criterion}
Nyquist and Shannon formulated the Nyquist zero-ISI criterion which serves as the basis of all modern digital communication systems. The pulse should satisfy the Nyquist criterion for the continuous-time waveform to be ISI-free. For a channel impulse response of $h(t)$, the condition for the received signal to be ISI free is given by,
\begin{equation}
h[n] = h(nT) = \delta_{n0} = \begin{cases} 
      1 & \text{ if } n=0 \\
      0 & \text{ if } n \neq 0 
\end{cases}
\end{equation}
where, the overall channel response $h[n]$ is given by the sampled sequence $(h_{TX}*h_{RX})(nT)$. If a pulse satisfies the above criterion, then that pulse is said to be Nyquist w.r.t. $T$. This condition translates to the following form in the frequency domain.
\begin{equation}
\frac{1}{T} \sum_{k=-\infty}^{\infty} H\left(f - \frac{k}{T} \right) = 1, \quad \forall f
\end{equation}
where, $H(f)$ is the CTFT of $h(t)$.

The Nyquist zero-ISI criterion puts a limit on the rate at which the transmit symbols can be transmitted. 
This rate $T$ is called the Nyquist rate. If the pulses are transmitted at this rate, there is no ISI and the adjacent pulses are orthogonal to each other. If the symbols are transmitted at a  higher  rate, adjacent pulses will not be orthogonal and this introduces ISI. This method of transmitting the symbols faster than $T$ is called the Faster-Than-Nyquist (FTN) Signalling.

\section{Faster-Than-Nyquist Signalling}
\label{sec:ftn_signalling}
The Nyquist zero-ISI criterion puts a limit on the rate at which the symbols can be transmitted. If the modulating pulse used is Nyquist w.r.t. $T$, then the transmit symbols cannot be packed closer than $T$ to ensure that there is no ISI in the received signal. In recent years, transmitting the symbols faster than $T$ while tolerating some ISI has attracted attention. This technique is called the Faster-Than-Nyquist (FTN) Signalling.

In FTN, pulse shaping is done with a pulse that is Nyquist w.r.t. $T$, but the symbols are spaced with a time period of $\tau T$ where $\tau < 1$. The factor $\tau$ is called the time acceleration factor. As the pulse used is Nyquist w.r.t. $ T$ and not $\tau T$, it violates the Nyquist zero-ISI criterion, and hence introduces ISI. This necessitates a complex transceiver structure that is capable of mitigating the ISI introduced by FTN signalling. The following chapter discusses FTN signalling in detail.


\chapter{FASTER-THAN-NYQUIST SIGNALLING}
\label{chap:ftn_intro}
The details of FTN signalling is discussed in this chapter. The ISI introduced by FTN is modelled mathematically in this chapter. In Section \ref{sec:sc_ftn}, the details and the formulation of expressions for received signal for single-carrier FTN system are discussed.  In Section \ref{sec:mc_ftn}, the details of multi-carrier FTN systems with OFDM modulation are discussed.

\section{Single-Carrier FTN System}
\label{sec:sc_ftn}
The FTN baseband single-carrier communication system is represented by the block diagram shown in Fig. \ref{fig:ftn_sc_bd_ftnintro}. 
\begin{figure}[h]
\hspace{-1cm}
{\scriptsize
\begin{tikzpicture}[auto, >=latex']
\draw
    node [input] (input1) {} 
    node [block, right of=input1, node distance=1.75cm] (mod) {Modulation}
    node [block, right of=mod, node distance=3.25cm] (tx_pulse) {$h_{TX}(t)$}
    node [sum, right of=tx_pulse, node distance=1.75cm] (add_noise) {+}
        node at (6.75, 1) [input] (input2) {}
    node [block, right of=add_noise, node distance=2cm] (matched_filter) {$h_{RX}(t)$}
    node [input, right of=matched_filter] (sampler) {}
    node [block, right of=sampler, node distance=2.5cm] (white) {Whitening}
    node [block, right of=white, node distance=2.5cm, text width = 1.5cm, align=center] (demod) {Equalize + Demod}
    node [output, right of=demod, node distance=1.75cm] (output1){}
    ;
    
    \sampler{0.75cm}{matched_filter}{sampler}
    \node at (10.5,-0.25) {rate $\tau T$};
    
    \draw[->](input1) -- node {$b[n]$}(mod);
    \draw[->] (mod) -- node {$s[n]$} node [below] {spaced $\tau T$} (tx_pulse);
    \draw[->](tx_pulse) -- node {}(add_noise);
    \draw[->](add_noise) -- node {$r(t)$}(matched_filter);
        \draw[->](input2) -- node {$\eta(t)$}(add_noise);
    \draw[-] (matched_filter) -- node {$y(t)$}(sampler);
    \draw[->](sampler) -- node {$y[n]$}(white);
    \draw[->](white) -- node {$z[n]$} (demod);
    \draw[->](demod) -- node {$\hat{b}[n]$} (output1);
\end{tikzpicture}}
\caption{Block diagram of a single carrier FTN communication system.}
\label{fig:ftn_sc_bd_ftnintro}
\end{figure}
At the transmitter, the modulated symbols are passed at a rate of $\tau T$ ($\tau < 1$ is called the time acceleration factor of the FTN system) through a pulse shaping filter $h_{TX}(t)$ (like SRRC) that is Nyquist with respect to a period $T$. The FTN waveform thus obtained is transmitted through an AWGN channel. At the receiver, the received waveform is passed through a filter $h_{RX}(t)$ that is matched to the transmit pulse $h_{TX}(t)$. The output of the filter is sampled at $\tau T$ to get discrete time samples. 

The transmit symbols are denoted by $s[n]$. The expression for the received signal $r(t)$ i.e. the input to the matched filter at the receiver is
\begin{equation}
r(t) = \sum_m s[m]h_{TX}(t-m\tau T) + \eta(t)
\end{equation}
where, $\eta(t)$ is the additive white Gaussian noise. Here, the transmitted symbols $s[n]$ are assumed to have unit average power and to be i.i.d. over $n$. The received signal is transmitted through the matched filter $h_{RX}(t)$. The output of the matched filter $y(t)$ is given by
\begin{equation}
y(t) = r(t)*h_{RX}(t) =  \sum_m s[m](h_{TX}*h_{RX})(t-m\tau T) + \eta'(t)
\end{equation}
where, $*$ represents convolution and $\eta'(t)$ is the noise coloured due to convolution with the matched filter. This is passed through the sampler at rate $\tau T$ and the output of this  sampler is
\begin{equation}
y[n] = y(n\tau T) = \sum_m s[m](h_{TX}*h_{RX})(n\tau T-m\tau T) + \eta'(n\tau T).
\end{equation}
Simplifying,
\begin{equation}
y[n] = \sum_m s[m](h_{TX}*h_{RX})[n-m] + \eta'[n]
\end{equation}
where, $(h_{TX}*h_{RX})[n]$ is the sampled sequence $(h_{TX}*h_{RX})(t)|_{\tau T}$ and $\eta'[n]$ is the sampled sequence $\eta'(t)|_{\tau T}$.

The overall system can now be represented as 
\begin{equation}
\label{eq:ftn_overall_eq}
    y[n] = h[n] * s[n] + \eta'[n]
\end{equation}
where, $h[n]$ is  the discrete sequence $(h_{TX}*h_{RX})(t)|_{\tau T}$. 

Note that the pulse $(h_{TX}*h_{RX})(t)$ is Nyquist w.r.t. $T$, and not w.r.t. $\tau T$. Hence, the sampled sequence  $h[n] = (h_{TX}*h_{RX})(t)|_{\tau T}$ is not Dirac delta. Hence, because of this faster transmission, the pulse introduces ISI to the system.

\begin{figure}[h]
    \centering
    \begin{subfigure}{0.45\textwidth}
        \centering
        \includegraphics[width=\textwidth]{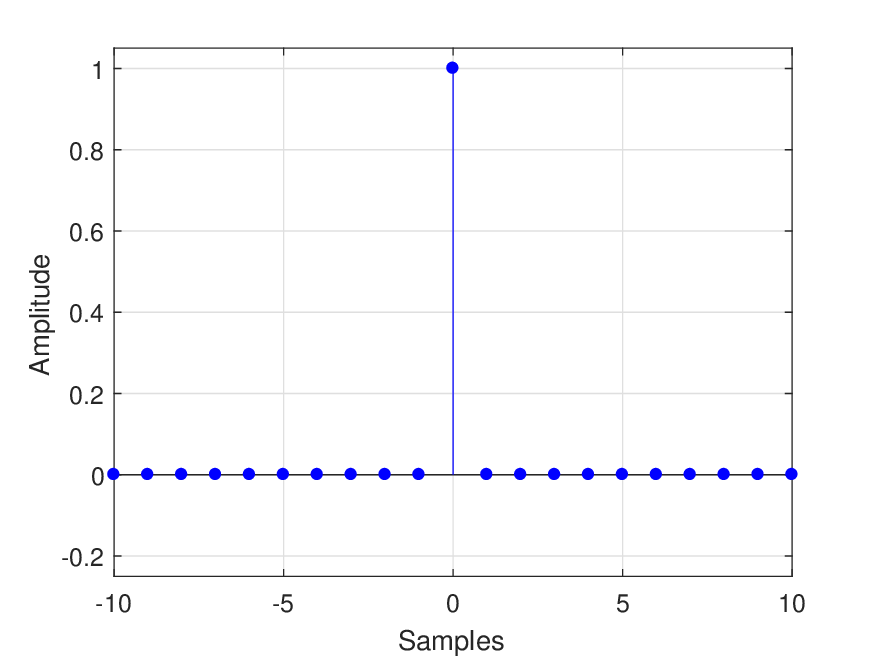}
        \caption{$\tau=1$}
    \end{subfigure}
    \begin{subfigure}{0.45\textwidth}
        \centering
        \includegraphics[width=\textwidth]{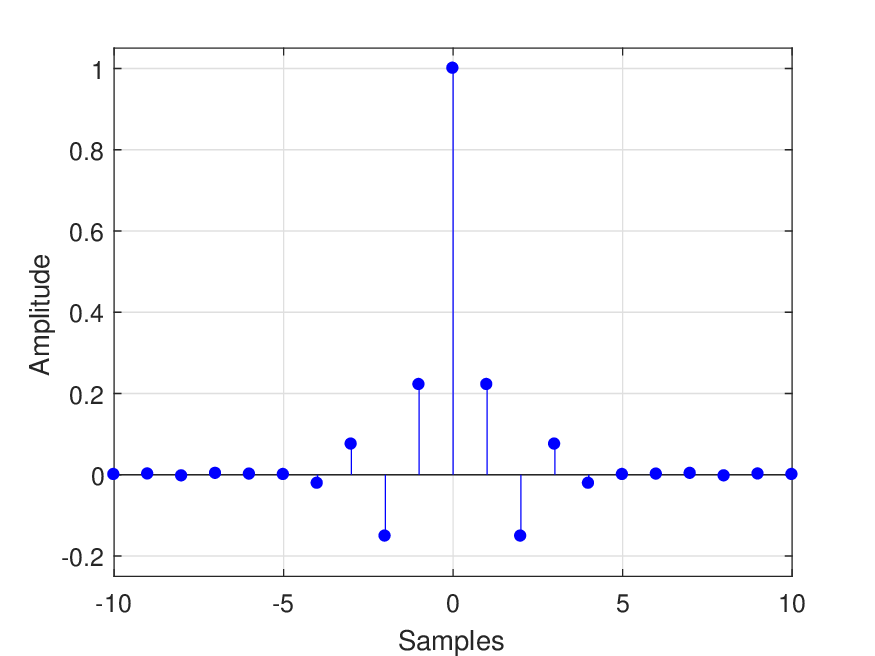}
        \caption{$\tau=0.8$}
    \end{subfigure}
    \begin{subfigure}{0.45\textwidth}
        \centering
        \includegraphics[width=\textwidth]{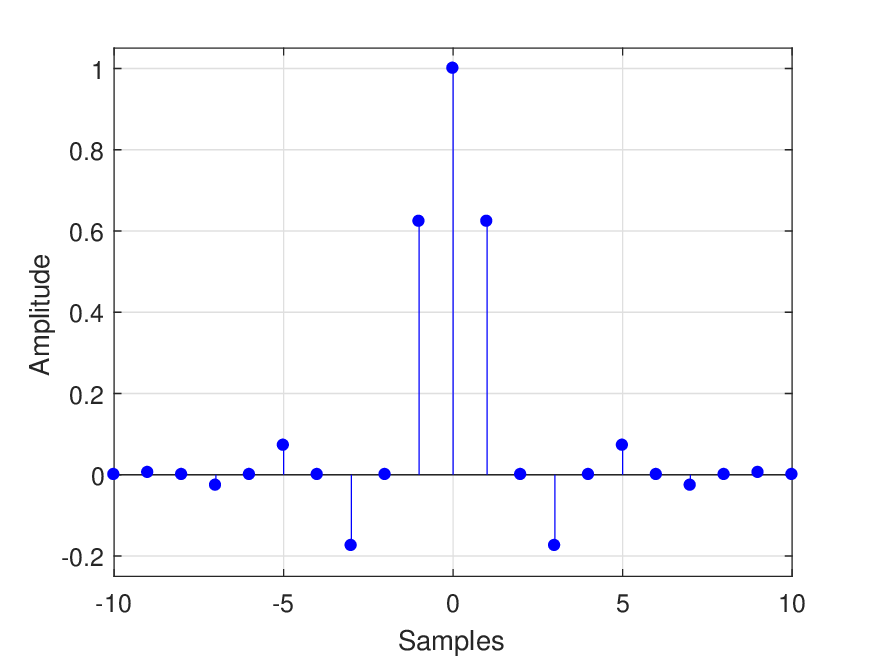}
        \caption{ $\tau=0.5$}
    \end{subfigure}
    \begin{subfigure}{0.45\textwidth}
        \centering
        \includegraphics[width=\textwidth]{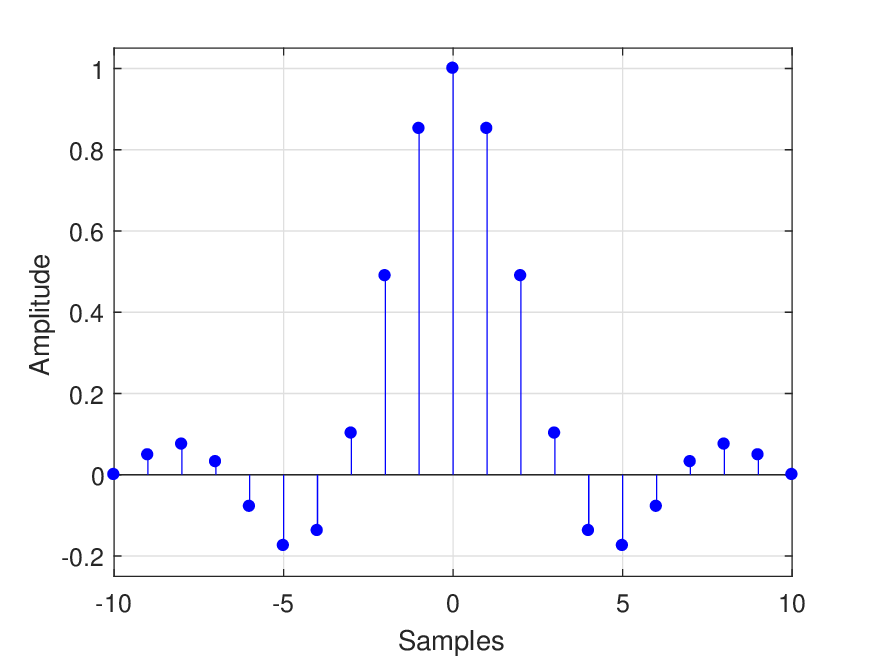}
        \caption{$\tau=0.3$}
    \end{subfigure}
    \caption{Discrete channel response $h[n]$ corresponding to the ISI introduced by FTN signalling with SRRC pulse of roll-off $\alpha=0.3$ for different values of time acceleration factor $\tau $.}
    \label{fig:ftn_isi_channel_response}
\end{figure}

Hence, from (\ref{eq:ftn_overall_eq}), it can be seen that the ISI introduced by FTN signalling can be modelled as convolution with an impulse response given by $h[n] = (h_{TX}*h_{RX})(n\tau T)$. The plots of $h[n]$ when SRRC pulse is used for modulation is given in Fig. \ref{fig:ftn_isi_channel_response}.
It can be seen that as the value of $\tau$ increases, both the magnitude and number of taps of ISI increase. As a result, FTN signalling leads to two opposing effects: as the value of $\tau$ decreases, the symbols are packed closer resulting in an increase in transmission rate. However, as the value of $\tau$ decreases, the intensity of ISI also increases resulting in an increase in the symbol error rate.

The overall equation of FTN in \ref{eq:ftn_overall_eq} can also be represented in the matrix form as
\begin{equation}
    \mathbf{y}=\mathbf{H}\mathbf{s} + \mathbf{\eta}'
\end{equation}
where, $\mathbf{s}$ and $\mathbf{y}$ are the vectors of transmitted and received sequences respectively, $\mathbf{\eta}'$ is the corresponding coloured noise and $\mathbf{H}$ is a large Toeplitz matrix formed by the channel $h[n]$. 

This ISI introduced by FTN needs to be equalized with the help of equalizers like MLSE. Using modulation schemes like OFDM and working with multi-carrier systems makes it much easier to mitigate this ISI. Hence, the FTN signalling in multi-carrier systems is discussed in the following section.

\section{Multi-Carrier FTN System}
\label{sec:mc_ftn}
The FTN baseband multi-carrier OFDM communication system is represented by the block diagram shown in  Fig. \ref{fig:ftn_mc_bd_ftnintro}.
\begin{figure}[h]
\hspace{-1.25cm}
{\scriptsize
\begin{tikzpicture}[auto, node distance=2.5cm, >=latex']
\draw
    node [input] (input1) {} 
    node [smallblock, right of=input1, node distance=1.5cm] (mod) {Mod}
    
    node [longblock, right of=mod, node distance=1.6cm, text width = 0.75em, align=center] (rx_sp) {\rotatebox{90}{S/P}}
    node [longblock, right of=rx_sp, node distance=1.15cm, text width = 0.75em, align=center] (ifft) {\rotatebox{90}{IFFT}}
    node [longblock, right of=ifft, node distance=0.65cm, text width = 0.75em, align=center] (rx_ps_cp) {\rotatebox{90}{P/S + CPA}}

    node [smallblock, right of=rx_ps_cp, node distance=1.7cm] (tx_pulse) {$h_{TX}(t)$}
    node [sum, right of=tx_pulse, node distance=1.4cm] (add_noise) {+}
        node at (8, 1.25) [input] (input2) {}
    node [smallblock, right of=add_noise, node distance=1.75cm] (matched_filter) {$h_{RX}(t)$}
    node [input, right of=matched_filter] (sampler) {}

    node [longblock, right of=sampler, node distance=0.35cm, text width = 0.75em, align=center] (tx_ps) {\rotatebox{90}{S/P + CPR}}
    node [longblock, right of=tx_ps, node distance=0.65cm, text width = 0.75em, align=center] (fft) {\rotatebox{90}{FFT}}
    node [longblock, right of=fft, node distance=1.15cm, text width = 0.75em, align=center] (tx_sp_cpr) {\rotatebox{90}{P/S}}     
            
    node [smallblock, right of=tx_sp_cpr, node distance=1.4cm, text width = 1.5cm, align=center] (demod) {Equalize + Demod}
    node [output, right of=demod, node distance=1.6cm] (output1){}
    ;
    
    \sampler{0.75cm}{matched_filter}{sampler}
    \node at (11.25,-0.25) {rate $\tau T$};
    
    \draw[->](input1) -- node {$b[n]$}(mod);
    
    \draw[->](mod) -- node {$s[n]$}(rx_sp);
    
    \draw[transform canvas={yshift=1.5em},->](rx_sp) -- node [below]{$\vdots$}(ifft);
    \draw[transform canvas={yshift=2.5em},->](rx_sp) -- node {$s[i]$}(ifft);
    \draw[transform canvas={yshift=-2.5em},->](rx_sp) -- node {}(ifft);
    \draw[transform canvas={yshift=-1.5em},->](rx_sp) -- node {}(ifft);
    
    \draw[transform canvas={yshift=1.5em},->](ifft) -- node [below]{$\vdots$}(rx_ps_cp); 
    \draw[transform canvas={yshift=2.5em},->](ifft) -- node {}(rx_ps_cp); 
    \draw[transform canvas={yshift=-2.5em},->](ifft) -- node {}(rx_ps_cp); 
    \draw[transform canvas={yshift=-1.5em},->](ifft) -- node {}(rx_ps_cp);        
    
    \draw[->] (rx_ps_cp) -- node [below] {$@ \tau T$} (tx_pulse);
    \draw[->](tx_pulse) -- node {}(add_noise);
    \draw[->](add_noise) -- node {$r(t)$}(matched_filter);
        \draw[->](input2) -- node {$\eta(t)$}(add_noise);
    \draw[-] (matched_filter) -- node {$y(t)$}(sampler);
    \draw[->](sampler) -- node {$y[n]$}(tx_ps);
    
    \draw[transform canvas={yshift=1.5em},->](tx_ps) -- node [below]{$\vdots$}(fft);
    \draw[transform canvas={yshift=2.5em},->](tx_ps) -- node {}(fft);
    \draw[transform canvas={yshift=-2.5em},->](tx_ps) -- node {}(fft);
    \draw[transform canvas={yshift=-1.5em},->](tx_ps) -- node {}(fft);
    
    \draw[transform canvas={yshift=1.5em},->](fft) -- node [below]{$\vdots$}(tx_sp_cpr); 
    \draw[transform canvas={yshift=2.5em},->](fft) -- node {$y[i]$}(tx_sp_cpr); 
    \draw[transform canvas={yshift=-2.5em},->](fft) -- node {}(tx_sp_cpr); 
    \draw[transform canvas={yshift=-1.5em},->](fft) -- node {}(tx_sp_cpr);

    \draw[->](tx_sp_cpr) -- node {} (demod);
    \draw[->](demod) -- node {$\hat{b}[n]$} (output1);
\end{tikzpicture}}
\caption{Block diagram of a multi-carrier OFDM FTN communication system. Here, CPA stands for cyclic prefix addition, CPR stands for cyclic prefix removal, S/P is serial-to-parallel, and P/S is parallel-to-serial.}
\label{fig:ftn_mc_bd_ftnintro}
\end{figure}
At the transmitter, after the N-point IFFT operation (where, N is the OFDM length) and parallel-to-serial conversion, cyclic prefix is added and the symbols are passed at a rate of $\tau T$ ($\tau < 1$) through a pulse shaping filter. The pulse shaping at the transmitter after IFFT has a response of $h_{TX}(t)$ that is Nyquist w.r.t. $T$. At the receiver, the matched filter has a response of $h_{RX}(t)$.

The overall system model of OFDM is given by
\begin{equation}
\label{eq:OFDM_model_main}
    y[i] = H[i]s[i] + \eta'[i] \quad i={0,\dots,N-1} 
\end{equation}
where, $H[i]$ is the $i$-th coefficient of N-point DFT of $(h_{TX}*h_{RX})(t)|_{\tau T}$ (i.e. $(h_{TX}*h_{RX})(t)$ sampled at $\tau T$) and $\eta'[i]$ is the coloured noise. It should be noted that each value of $i$ corresponds to each sub-carrier of OFDM. 

\begin{figure}[h]
    \centering
    \begin{subfigure}{0.45\textwidth}
        \centering
        \includegraphics[width=\textwidth]{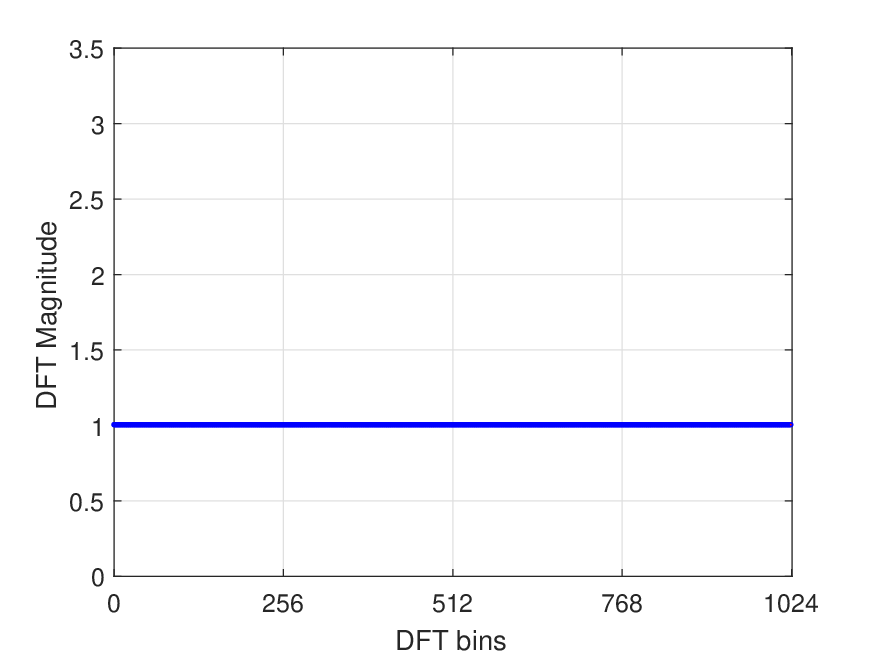}
        \caption{$\tau=1$}
    \end{subfigure}
    \begin{subfigure}{0.45\textwidth}
        \centering
        \includegraphics[width=\textwidth]{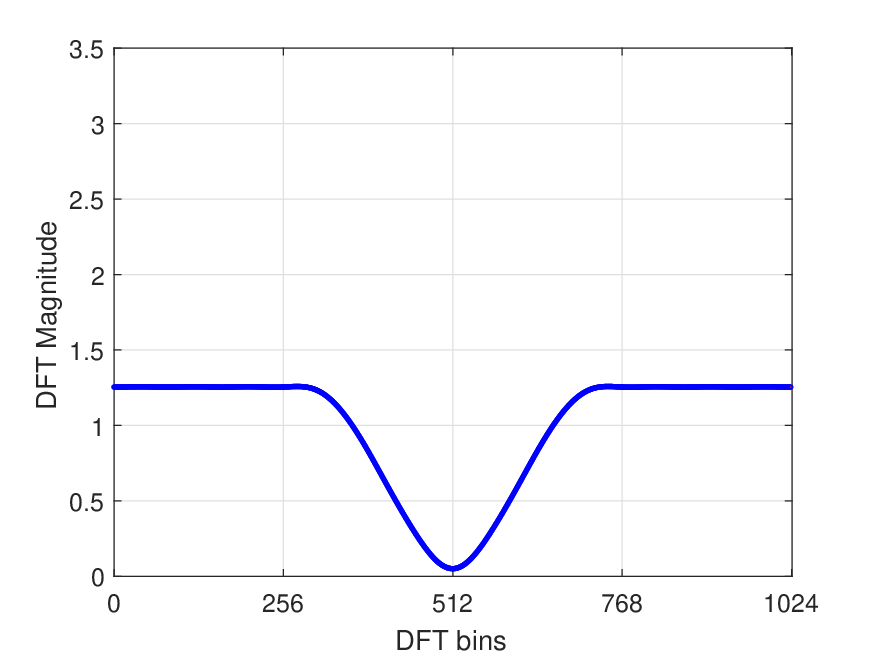}
        \caption{$\tau=0.8$}
    \end{subfigure}
    \begin{subfigure}{0.45\textwidth}
        \centering
        \includegraphics[width=\textwidth]{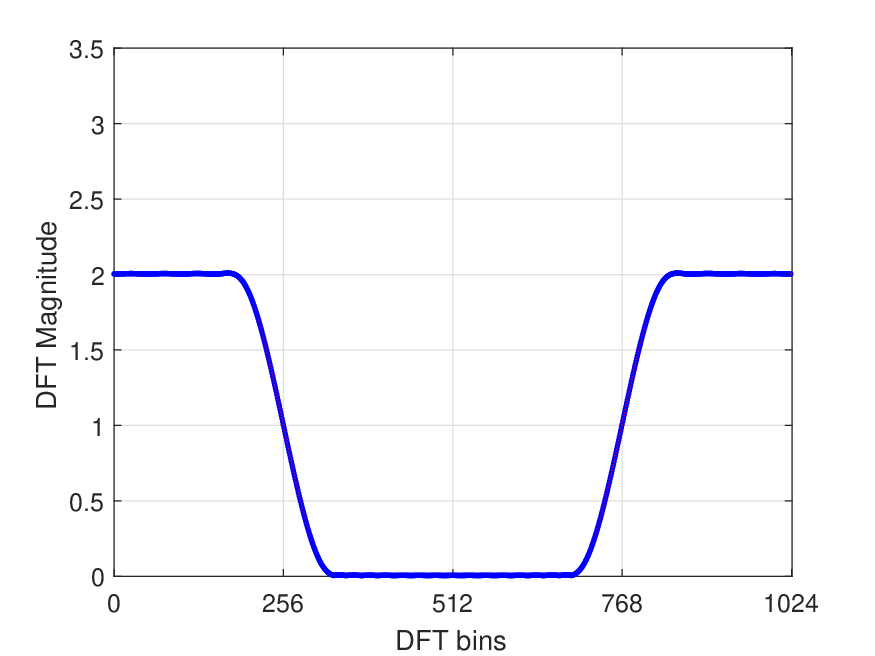}
        \caption{ $\tau=0.5$}
    \end{subfigure}
    \begin{subfigure}{0.45\textwidth}
        \centering
        \includegraphics[width=\textwidth]{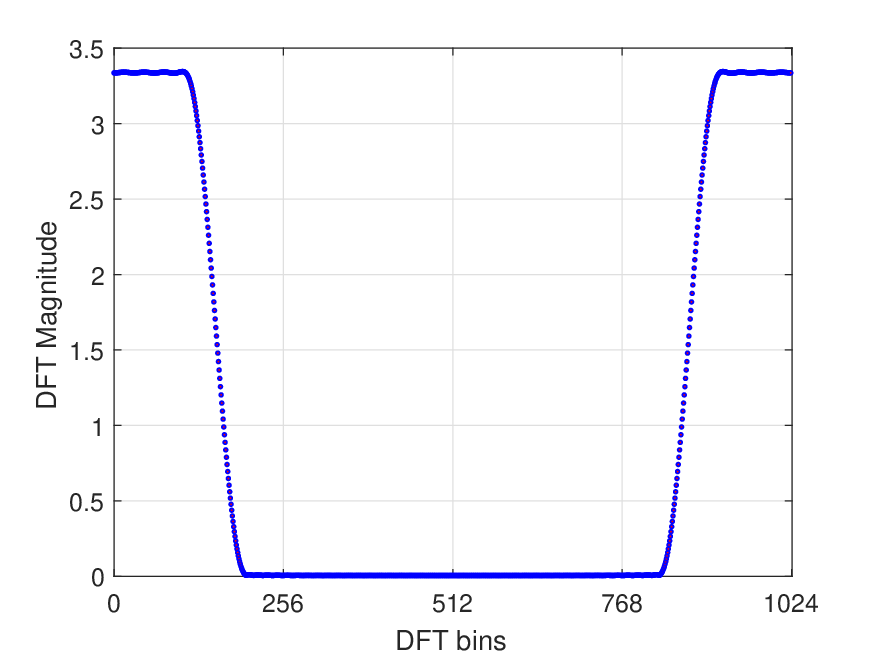}
        \caption{$\tau=0.3$}
    \end{subfigure}
    \caption{The magnitude $|H[i]|$ of the 1024-point DFT coefficients of $h[n]$ corresponding to the ISI introduced by FTN signalling with SRRC pulse of roll-off $\alpha=0.3$ for different values of time acceleration factor $\tau $.}
    \label{fig:ftn_isi_freq_response}
\end{figure}

In equation (\ref{eq:OFDM_model_main}), each value of $i$ corresponds to a sub-carrier of the multi-carrier system. The transmitted symbols in each sub-carrier $i$ is scaled by $H[i]$. Here, $H[i]$ is the $i$-th coefficient of N-point DFT of the overall response corresponding to the ISI introduced by FTN signalling. This scaling can be easily equalized with the help of an one tap equalizer as shown in (\ref{eq:one_tap_eq}). Because of this, handling the ISI introduced by FTN signalling becomes much easier in OFDM FTN systems.\vspace{-0.75cm}
\begin{equation}
\label{eq:one_tap_eq}
\begin{split}
\hat{y}[i] & = y[i]/H[i] \\
& = s[i] + \eta'[i]/H[i]; \quad \text{if } H[i] \neq0,  \quad i={0,\dots,N-1}
\end{split}
\end{equation}
where, $\hat{y}[i]$ is the estimated symbol in $i$-th sub-carrier. It can be noted here that, for all $i$ where $|H[i]| < 1$, the noise value $ \eta'[i]$ is scaled up. This leads to a degradation in the error rate performance of this sub-carrier. The variation of $|H[i]|$, the $i$-th coefficient of N-point DFT of $h_{RC}(t)|_{\tau T}$, as a function of $i$ is shown in Fig. \ref{fig:ftn_isi_freq_response} taking the modulating pulse as SRRC. 
It can be seen that the magnitude of $H[i]$ is highest for the sub-carriers at the sides. For the sub-carriers in the middle, the magnitude is low. Because of this, the error performance of the sub-carriers at the corners is better than that of the ones in the middle. In order to alleviate this, the power allotted to sub-carriers can be varied depending on the value of $H[i]$ for that sub-carrier. Further, different modulation schemes can be used in different sub-carriers. Chapter \ref{chap:waterfilling_loading} discusses these techniques, which ultimately leads to better transmission rates of FTN systems. The following chapters discuss the the existing literature on the field of FTN and formulate the focus of this thesis.

\chapter{LITERATURE SURVEY}
\label{chap:literature_survey}
Some of the early results in the field of FTN signalling were published by Mazo. \citep{6772210,21281} derive the limits on the FTN time acceleration factor $\tau_{Mazo}$ such that the minimum distance between the transmitted symbols in the modified constellation is unchanged. This result implied that if FTN signalling does not decreases the minimum distance in the waveforms, then the probability of bit error does not increase. 

Tufts derived an analytical framework for FTN signalling with an MMSE equalizer and ended the previous claims that the data cannot be transmitted at a rate faster than the Nyquist rate. \citep{1445612} shows that it is indeed possible to send data faster than Nyquist rate in short busts. \citep{1054187} also shows that transmitting data faster than Nyquist is possible. This paper follows a slightly different approach of looking at FTN, instead of transmitting faster in time domain, this method reduces the system bandwidth slightly below the Nyquist bandwidth. 

\citep{1450960,1054829} revisit the the framework proposed by Viterbi and use the Viterbi decoder to handle the ISI introduced by multi-tap channels.  \citep{6771898} studies the feasibility of FTN with certain modulation schemes like binary and QAM modulation. However, these results were limited by the computational capacities of that time as the heavy ISI introduced by FTN leads to a very high decoder complexity for larger modulation schemes.  \citep{380028} shows significant improvements in the transmit filter for FTN using a whitening and matched filter at the receiver.

In recent years, significant work is done by John B. Anderson and Fredrik Rusek. \citep{4777625,4150682} show that the capacity of FTN systems for a finite alphabet is significantly higher than that of systems that use orthogonal signalling schemes.  \citep{6479673} is a summary of key results in FTN signalling.  

 \citep{5539725} examines the asymptotic optimality of binary FTN signalling. This paper shows that the capacity of a system can be achieved by using a transmit pulse which results in the same PSD as given by solving the water-filling optimization problem.  \citep{8320347} proposes ideas like MMSE channel shortening to  counter the issue of large number of ISI taps in severe FTN systems. It demonstrates a trade off between  the receiver complexity and performance of the system.

  \citep{8356261} proposes automatic trellis generation to design the equalizers for FTN systems.   \citep{7476106} discusses the performance of  non-orthogonal OFDM transceivers. This method is equivalent to performing Faster than Nyquist signalling in the frequency domain.

 \citep{7774821} proposes the use of Gaussian pulses and extended Gaussian functions in FTN systems to get a performance gain over the  traditional SRRC pulse.  \citep{7284519,8322513} discuss different types of liner precoding techniques such as Singular Value Decomposition (SVD), G-to-Minus-Half (GTMH) and Cholesky Decomposition. This paper shows that the Cholesky Decomposition and GTMH precoding performs better than the SVD precoding at the cost of broadening the signal spectrum. Some improvements on GTMH precoding were also proposed in   \citep{8401480}.

This thesis builds upon these works, analyses the pulse shapes closely, and studies the effect of ISI introduced by FTN signalling. The modulating pulse shapes are analysed and the constraints on pulse shapes are derived so that the performance can match that of a Nyquist system.
The major focus is on analysing the capacity of FTN systems as a function of the time acceleration factor. Techniques like power allocation and adaptive bit loading are proposed to increase the throughput of OFDM FTN systems.


\chapter{PROBLEM FORMULATION AND LAYOUT}
\label{chap:problem_formulation}
In Chapter \ref{chap:introduction}, the basics of a digital communication system was introduced along with the elementary concept of Nyquist ISI criterion. The idea of FTN signalling was motivated from the Nyquist ISI criterion, which dictates a limit on the rate at which symbols can be transmitted. It was shown that the rate of transmission could be increased by ignoring the Nyquist ISI criterion and transmitting at a rate Faster Than the Nyquist (FTN) rate. However, it was also shown that the FTN signalling increases the intensity of ISI. As a result, there is an interplay of two opposing effects in an FTN system: the closer packing of symbols improves the transmission rate, where as the ISI worsens the error rate performance.

In the first part of this thesis, the ISI introduced by FTN signalling is examined in detail. Conditions are derived so that the ISI can be inverted completely and symbols can be detected in a symbol-by-symbol manner. The ISI introduced by FTN signalling was modelled as a multi-tap channel in Chapter \ref{chap:ftn_intro}. As shown in (\ref{eq:ftn_overall_eq}), this response depends on the modulating pulse and the time acceleration factor $\tau$. In Chapter \ref{chap:isi_perspective}, the condition on the pulse shape and the time acceleration factor is derived so that the ISI can be mitigated completely.

In the second part of the thesis, the transmission rates of FTN systems are examined. The theoretical limits on the highest achievable transmission rates i.e. the capacity of FTN systems are calculated. In Chapter \ref{chap:capacity_perspective}, the capacity of FTN system as a function of its time acceleration factor is calculated. The behaviour of this capacity as a function of time acceleration factor is examined. It is also argued how the behaviour of capacity as a function of time acceleration factor reinforces the conditions derived in Chapter \ref{chap:isi_perspective} for complete ISI mitigation.

Finally, several techniques are discussed in Chapter \ref{chap:waterfilling_loading} that help soften the effect of ISI on error rate and hence increase the throughput of the system. These techniques are implemented and the simulation results are shown in  Chapter \ref{chap:simulation}.


\chapter{CONDITIONS FOR COMPLETE ISI MITIGATION}
\label{chap:isi_perspective}
The ISI introduced by an FTN system depends on the modulating pulse and the time acceleration factor $\tau$. Under certain conditions on pulse shapes and $\tau$, the ISI can be avoided completely leading to a symbol-by-symbol detection. In Section \ref{sec:analysis_sc} of this chapter, these conditions are derived for single carrier FTN systems. In Section \ref{sec:analysis_ofdm}, these conditions are derived for multi carrier OFDM FTN systems.

\section{Pulse Shape Constraint in Single-Carrier FTN Systems}
\label{sec:analysis_sc}
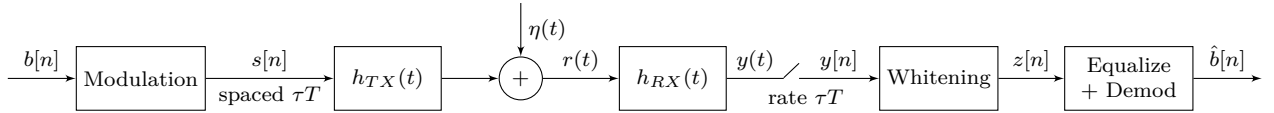
\begin{figure}[h]
\hspace{-1cm}
{\scriptsize
\begin{tikzpicture}[auto, >=latex']
\draw
    node [input] (input1) {} 
    node [block, right of=input1, node distance=1.75cm] (mod) {Modulation}
    node [block, right of=mod, node distance=3.25cm] (tx_pulse) {$h_{TX}(t)$}
    node [sum, right of=tx_pulse, node distance=1.75cm] (add_noise) {+}
        node at (6.75, 1) [input] (input2) {}
    node [block, right of=add_noise, node distance=2cm] (matched_filter) {$h_{RX}(t)$}
    node [input, right of=matched_filter] (sampler) {}
    node [block, right of=sampler, node distance=2.5cm] (white) {Whitening}
    node [block, right of=white, node distance=2.5cm, text width = 1.5cm, align=center] (demod) {Equalize + Demod}
    node [output, right of=demod, node distance=1.75cm] (output1){}
    ;
    
    \sampler{0.75cm}{matched_filter}{sampler}
    \node at (10.5,-0.25) {rate $\tau T$};
    
    \draw[->](input1) -- node {$b[n]$}(mod);
    \draw[->] (mod) -- node {$s[n]$} node [below] {spaced $\tau T$} (tx_pulse);
    \draw[->](tx_pulse) -- node {}(add_noise);
    \draw[->](add_noise) -- node {$r(t)$}(matched_filter);
        \draw[->](input2) -- node {$\eta(t)$}(add_noise);
    \draw[-] (matched_filter) -- node {$y(t)$}(sampler);
    \draw[->](sampler) -- node {$y[n]$}(white);
    \draw[->](white) -- node {$z[n]$} (demod);
    \draw[->](demod) -- node {$\hat{b}[n]$} (output1);
\end{tikzpicture}}
\caption{Block diagram of a single carrier FTN communication system.}
\label{fig:SC_BD_channinv_chap}
\end{figure}

As discussed in Chapter \ref{chap:ftn_intro}, the FTN baseband single-carrier communication system is represented by the block diagram shown in Fig. \ref{fig:SC_BD_channinv_chap}. 
At the transmitter, the modulated symbols are passed at a rate of $\tau T$ ($\tau < 1$) through a pulse shaping filter $h_{TX}(t)$ that is Nyquist with respect to $T$. The FTN waveform thus obtained is transmitted through an AWGN channel. At the receiver, the received waveform is passed through a filter $h_{RX}(t)$ that is matched to the transmit pulse $h_{TX}(t)$. The output of the filter is sampled at $\tau T$ to get discrete time samples.

As discussed in Chapter \ref{chap:ftn_intro}, the overall system can be represented as 
\begin{equation}
    y[n] = h[n] * s[n] + \eta'[n]
\end{equation}
where, $h[n]$ is  the discrete sequence $(h_{TX}*h_{RX})(t)|_{\tau T}$, $\eta'$ is the coloured noise due to the FTN signalling and $*$ represents convolution. This can be represented in the matrix form as
\begin{equation}
\label{eq:system_model}
    \mathbf{y}=\mathbf{H}\mathbf{s} + \mathbf{\eta}'
\end{equation}
where, $\mathbf{s}$ and $\mathbf{y}$ are the vectors of transmitted and received sequences respectively, $\mathbf{\eta}'$ is the corresponding coloured noise and $\mathbf{H}$ is a large Toeplitz matrix formed by the channel $h[n]$. 

One of the widely used transmit pulse in digital communication systems is the SRRC pulse. Rest of the analysis in this chapter is done considering the modulating pulse to be SRRC.

If the modulating pulse is SRRC, $h[n]$ is replaced by
\begin{equation}
h[n] = (h_{RX}*h_{TX})(n\tau T) = (h_{SRRC}*h_{SRRC})(n\tau T) = h_{RC}(n\tau T).
\end{equation}

Let $h[n]$ in this particular case be denoted by $h_{RC}[n]$. Hence, the matrix equation in (\ref{eq:system_model}) becomes
\begin{equation}
\label{eq:rc_system_model}
    \mathbf{y}=\mathbf{H_{RC}}\mathbf{s} + \mathbf{\eta}'
\end{equation}
where, $\mathbf{H_{RC}}$ is a large Toeplitz matrix formed by the channel $h_{RC}[n]$.

To estimate the transmit sequence $\hat{\mathbf{s}}$ symbol-by-symbol from received sequence ${\mathbf{y}}$, $\mathbf{H_{RC}}$ should be invertible. It is invertible when its eigenvalues are non-zero.
Since $\mathbf{H_{RC}}$ is a large Toeplitz matrix, its eigenvalues are approximately given by the DFT coefficients of $h_{RC}[n]$ \citep{toeplitz}. Hence, the analysis of DFT coefficients of $h_{RC}[n]$ is done below. 

As $h_{RC}[n]$ is the RC pulse sampled at $\tau T$, its DTFT ${H_{RC}(e^{j\omega})} $ is given by
\begin{equation}
    \label{eq:RC_DFT}
    {H_{RC}(e^{j\omega})} = \frac{1}{\tau T}\sum_{k=-\infty}^\infty \mathcal{H}_{RC}\left(\frac{\omega-2\pi k}{2\pi \tau T}\right)
\end{equation}
where, $\mathcal{H}_{RC}(f)$ is the CTFT of the RC pulse. Note that the CTFT $\mathcal{H}_{RC}(f)$ of RC pulse is bandlimited by $(1+\alpha)/2T$ where, $\alpha$ is the roll-off factor i.e. $\mathcal{H}_{RC}(f) = 0$ for $|f|>(1+\alpha)/2T$.

\begin{figure}[h]
    \centering
    \includegraphics[width=\textwidth]{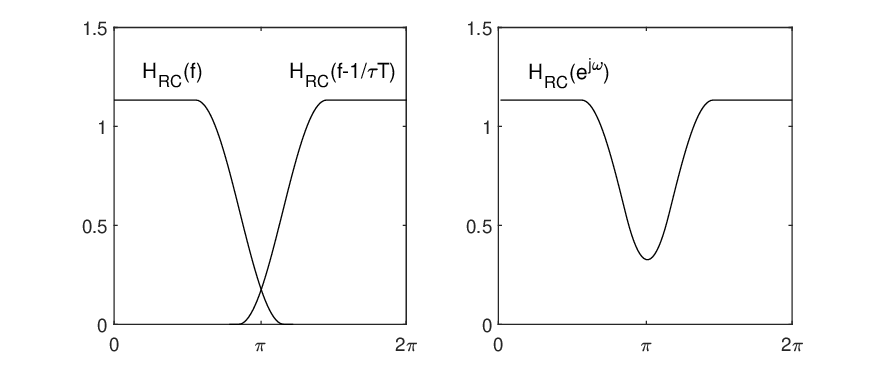}
    \caption{The DTFT ${H_{RC}(e^{j\omega})}$ when the adjacent copies of the CTFT $\mathcal{H}_{RC}(f)$  are overlapping. In this case, the overall response is non-zero for all frequencies.}
    \label{fig:SC_channel_yesop}
\end{figure}

If the DTFT ${H_{RC}(e^{j\omega})}$ is non-zero for all frequencies $\omega$, the DFT coefficients are non-zero. To satisfy this condition, the passbands or the transition bands of adjacent copies should overlap as shown in Fig. \ref{fig:SC_channel_yesop}. In this case, the overall response ${H_{RC}(e^{j\omega})}$ is non-zero for all frequencies $\omega$ i.e. ${H_{RC}(e^{j\omega})} > 0$, $\forall \omega$.

\begin{figure}[h]
    \centering
    \includegraphics[width=\textwidth]{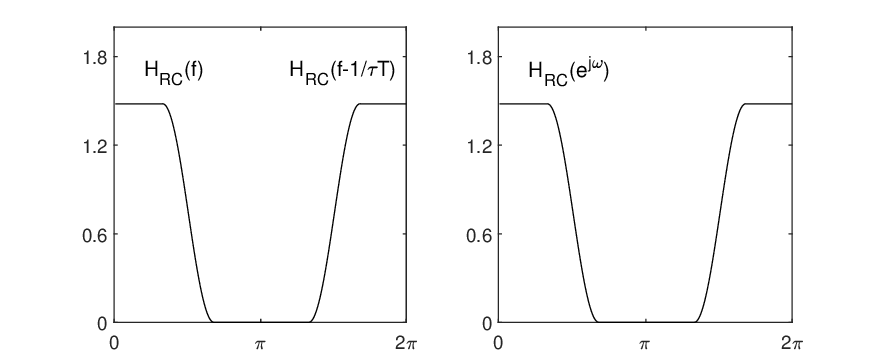}
    \caption{The DTFT ${H_{RC}(e^{j\omega})}$ when the adjacent copies of the CTFT $\mathcal{H}_{RC}(f)$  are non-overlapping.  In this case, there are certain frequencies $\omega$ where the overall response is zero.}
    \label{fig:SC_channel_noop}
\end{figure}

If the values of $\tau$ are very small, the adjacent copies are non-overlapping as shown in Fig. \ref{fig:SC_channel_noop}. In this case, there are certain frequencies $\omega$ where the overall response is zero i.e. ${H_{RC}(e^{j\omega})} = 0 $, for some $\omega$.

In order to ensure that the adjacent copies are overlapping, $\tau$ should satisfy the following condition.
\begin{equation*}
    \frac{1+\alpha}{2T} > \frac{1}{\tau T} - \frac{1+\alpha}{2T}.
\end{equation*}
This condition arrives from the facts that the adjacent copies are separated by $1/\tau T$ and both copies are bandlimited between $-(1+\alpha)/2T$ and $(1+\alpha)/2T$ around their centre. Simplifying,
\begin{equation}
    \label{eq:condition_single} (1+\alpha)\tau>1.
\end{equation}

If the condition in (\ref{eq:condition_single}) is satisfied, the DFT coefficients of $h_{RC}[n]$ are non-zero. This implies that the matrix $\mathbf{H_{RC}}$ in (\ref{eq:rc_system_model}) has all non-zero eigenvalues and hence is invertible. So, the ISI introduced by FTN can be mitigated completely and symbol-by-symbol detection is possible. One way to do this is by using precoders and postcoders that are designed by taking advantage of the invertibility of the matrix $\mathbf{H_{RC}}$. This is discussed in detail in our published work \citep{8904284}. 

\section{Pulse Shape Constraint in  Multi-Carrier OFDM FTN Systems}\label{sec:analysis_ofdm}
In the previous section, the condition on pulse shaping for symbol-by-symbol detection was derived considering single-carrier FTN systems. Same condition can also be derived in an easier way considering OFDM FTN systems.

As discussed in Chapter \ref{chap:ftn_intro}, the FTN baseband multi-carrier OFDM communication system is represented by the block diagram shown in  Fig. \ref{fig:ftn_mc_bd_constraint}
\begin{figure}[h]
\hspace{-1.25cm}
\vspace{-12pt}
{\scriptsize
\begin{tikzpicture}[auto, node distance=2.5cm, >=latex']
\draw
    node [input] (input1) {} 
    node [smallblock, right of=input1, node distance=1.5cm] (mod) {Mod}
    
    node [longblock, right of=mod, node distance=1.6cm, text width = 0.75em, align=center] (rx_sp) {\rotatebox{90}{S/P}}
    node [longblock, right of=rx_sp, node distance=1.15cm, text width = 0.75em, align=center] (ifft) {\rotatebox{90}{IFFT}}
    node [longblock, right of=ifft, node distance=0.65cm, text width = 0.75em, align=center] (rx_ps_cp) {\rotatebox{90}{P/S + CPA}}

    node [smallblock, right of=rx_ps_cp, node distance=1.7cm] (tx_pulse) {$h_{TX}(t)$}
    node [sum, right of=tx_pulse, node distance=1.4cm] (add_noise) {+}
        node at (8, 1.25) [input] (input2) {}
    node [smallblock, right of=add_noise, node distance=1.75cm] (matched_filter) {$h_{RX}(t)$}
    node [input, right of=matched_filter] (sampler) {}

    node [longblock, right of=sampler, node distance=0.35cm, text width = 0.75em, align=center] (tx_ps) {\rotatebox{90}{S/P + CPR}}
    node [longblock, right of=tx_ps, node distance=0.65cm, text width = 0.75em, align=center] (fft) {\rotatebox{90}{FFT}}
    node [longblock, right of=fft, node distance=1.15cm, text width = 0.75em, align=center] (tx_sp_cpr) {\rotatebox{90}{P/S}}     
            
    node [smallblock, right of=tx_sp_cpr, node distance=1.4cm, text width = 1.5cm, align=center] (demod) {Equalize + Demod}
    node [output, right of=demod, node distance=1.6cm] (output1){}
    ;
    
    \sampler{0.75cm}{matched_filter}{sampler}
    \node at (11.25,-0.25) {rate $\tau T$};
    
    \draw[->](input1) -- node {$b[n]$}(mod);
    
    \draw[->](mod) -- node {$s[n]$}(rx_sp);
    
    \draw[transform canvas={yshift=1.5em},->](rx_sp) -- node [below]{$\vdots$}(ifft);
    \draw[transform canvas={yshift=2.5em},->](rx_sp) -- node {$s[i]$}(ifft);
    \draw[transform canvas={yshift=-2.5em},->](rx_sp) -- node {}(ifft);
    \draw[transform canvas={yshift=-1.5em},->](rx_sp) -- node {}(ifft);
    
    \draw[transform canvas={yshift=1.5em},->](ifft) -- node [below]{$\vdots$}(rx_ps_cp); 
    \draw[transform canvas={yshift=2.5em},->](ifft) -- node {}(rx_ps_cp); 
    \draw[transform canvas={yshift=-2.5em},->](ifft) -- node {}(rx_ps_cp); 
    \draw[transform canvas={yshift=-1.5em},->](ifft) -- node {}(rx_ps_cp);        
    
    \draw[->] (rx_ps_cp) -- node [below] {$@ \tau T$} (tx_pulse);
    \draw[->](tx_pulse) -- node {}(add_noise);
    \draw[->](add_noise) -- node {$r(t)$}(matched_filter);
        \draw[->](input2) -- node {$\eta(t)$}(add_noise);
    \draw[-] (matched_filter) -- node {$y(t)$}(sampler);
    \draw[->](sampler) -- node {$y[n]$}(tx_ps);
    
    \draw[transform canvas={yshift=1.5em},->](tx_ps) -- node [below]{$\vdots$}(fft);
    \draw[transform canvas={yshift=2.5em},->](tx_ps) -- node {}(fft);
    \draw[transform canvas={yshift=-2.5em},->](tx_ps) -- node {}(fft);
    \draw[transform canvas={yshift=-1.5em},->](tx_ps) -- node {}(fft);
    
    \draw[transform canvas={yshift=1.5em},->](fft) -- node [below]{$\vdots$}(tx_sp_cpr); 
    \draw[transform canvas={yshift=2.5em},->](fft) -- node {$y[i]$}(tx_sp_cpr); 
    \draw[transform canvas={yshift=-2.5em},->](fft) -- node {}(tx_sp_cpr); 
    \draw[transform canvas={yshift=-1.5em},->](fft) -- node {}(tx_sp_cpr);

    \draw[->](tx_sp_cpr) -- node {} (demod);
    \draw[->](demod) -- node {$\hat{b}[n]$} (output1);
\end{tikzpicture}}
\caption{Block diagram of a multi carrier OFDM FTN communication system. Here, CPA stands for cyclic prefix addition, CPR stands for cyclic prefix removal, S/P is serial-to-parallel, and P/S is parallel-to-serial.}
\label{fig:ftn_mc_bd_constraint}
\end{figure}
\vspace{-8pt}

At the transmitter, after the N-point IFFT operation and parallel-to-serial conversion, cyclic prefix is added and the symbols are passed at a rate of $\tau T$ ($\tau < 1$) through a pulse shaping filter. The pulse shaping at the transmitter after IFFT has a response of $h_{TX}(t)$. At the receiver, the matched filter has a response of $h_{RX}(t)$.

The overall system model of OFDM is given by
\begin{equation}
\label{eq:OFDM_model}
    y[i] = H[i]s[i] + \eta'[i] \quad i={0,\dots,N-1} 
\end{equation}
where $H[i]$ is the $i$-th coefficient of N-point DFT of $(h_{TX}*h_{RX})(t)|_{\tau T}$ (i.e. $(h_{TX}*h_{RX})(t)$ sampled at $\tau T$) and $\eta'[i]$ is the coloured noise. It should be noted that each value of $i$ corresponds to each sub-carrier of OFDM. 

To estimate $s[i]$ from $y[i]$, none of the $H[i]$ should be equal to zero i.e. none of the DFT coefficients of $(h_{TX}*h_{RX})(t)|_{\tau T}$ should be zero. In the particular case of SRRC modulating pulse, none of the DFT coefficients of $h_{RC}(t)|_{\tau T}$ should be zero. Proceeding further, following similar methods as in the previous section, the condition on pulse can be derived for OFDM FTN system and is seen to same as that for the single-carrier FTN system in (\ref{eq:condition_single}) i.e.
\begin{equation}
\label{eq:condition_single_again}
 (1+\alpha)\tau>1.
\end{equation}

If the condition (\ref{eq:condition_single_again}) is not satisfied, then there exists some frequencies $\omega$ where $H_{RC}(e^{j\omega})$ goes to zero (see Fig. \ref{fig:SC_channel_noop} for example). The sub-carriers corresponding to those frequencies have $H[i]=0$. Thus, those sub-carriers cannot be used to transmit symbols. On the contrary, if the condition (\ref{eq:condition_single_again}) is satisfied, the DTFT $H_{RC}(e^{j\omega})$ is non-zero for all frequencies $\omega$. Hence, all the sub-carriers can be used to transmit symbols. Precoding can be used in this case to utilize all the sub-carriers and reach the BER performance of a Nyquist system. This is discussed in detail in our published work \citep{8904284}.

Now that the conditions are derived to handle the ISI introduced by FTN signalling, the next chapter discusses the maximum transmission rates that can be achieved with FTN signalling.


\chapter{CAPACITY OF FTN SYSTEMS}
\label{chap:capacity_perspective}
In the previous chapter, conditions were derived on the pulse shape so that the detection can be done in a symbol-by-symbol manner. Irrespective of this ISI or the symbol-by-symbol detection, the transmission rates of FTN systems need to be examined in order to see if FTN signalling is really more useful than Nyquist signalling. In this chapter, the transmission rates of FTN systems are examined theoretically. The maximum possible transmission rate, the capacity, of FTN system is derived in this chapter.

In Section \ref{sec:ct_cap} of this chapter, the capacity of a generic continuous time communication system is explained. In Section \ref{sec:dt_cap}, the capacity of a discrete time communication system is explained. Finally, in Section \ref{sec:ftn_cap}, the capacity of FTN system is derived and analysis of this capacity for modulation pulses like SRRC and rectangular pulses is done.

\section{Capacity of Continuous-Time Systems}
\label{sec:ct_cap}
As discussed in Chapter \ref{chap:introduction}, in a baseband digital communication system, the expression for the received signal $r(t)$ when the transmit symbols $s[n]$ are linearly modulated and transmitted over an AWGN channel is 
\begin{equation}
r(t) = \sum_m s[m]h_{TX}(t-mT) + \eta(t)
\end{equation}
where, $h_{TX}(t)$ is the modulating pulse used at the transmitter, $\eta(t)$ is the additive white Gaussian noise, and $T$ is the time period with which the symbols are spaced. Here, the transmitted symbols $s[n]$ are assumed to have unit average power and to be i.i.d. over $n$.

Shannon gave the theoretical limit on the maximum transmission rate achievable over an AWGN channel of the above form. That limit, called the channel capacity, depends on the magnitude response of the modulating pulse $h_{TX}(t)$ and the average SNR of the received signal. The following subsections discuss the expression for channel capacity for two cases when the modulating pulse is sinc in time domain and the modulating pulse is not sinc in time domain.
	
\subsection{Capacity of systems that use sinc modulating pulses}
When the modulating pulse $h_{TX}(t)$ is sinc in time domain, the PSD of the transmitted signal is flat over the signal bandwidth. A simple proof of this is given in Appendix \ref{sec:append_psd} that uses the assumption that the transmitted symbols $s[n]$ have unit average power and are i.i.d. over $n$. When the PSD of the transmitted signal is flat over the utilized bandwidth, the channel capacity for a given average SNR is given by
\begin{equation}
    C_{flat} = W\log_2 (1+\text{SNR}) \text{ bits/s}
    \label{eq:c_flat}
\end{equation}
where, $W$ is the bandwidth of the signal.

The above expression for $C_{flat}$ is valid only when the modulating pulse $h_{TX}(t)$ is sinc in time domain or flat in frequency domain. Next subsection discusses the capacity of the system when the modulating pulse is not sinc in time domain.

\subsection{Capacity of systems that use non-sinc modulating pulses}
In the previous subsection, the channel capacity expression was given assuming that the pulse $h_{TX}(t)$ was a sinc pulse. However, sinc pulses are not used in practical systems. Instead, other pulses that satisfy the Nyquist ISI criterion are used. The expression for capacity $C_{flat}$ in (\ref{eq:c_flat}) cannot be used to calculate the capacity of such systems. 

In order to obtain the new capacity expression in this case, the PSD of the transmitted signal needs to be calculated. As the transmitted symbols $s[n]$ are assumed to have unit average power and to be i.i.d. over $n$, the PSD of the transmitted signal is equal to the squared magnitude response $W|H(f)|^2$ of the modulating pulse $h_{TX}(t)$. A simple proof of this is given in Appendix \ref{sec:append_psd}. 

Now that the PSD of the transmitted signal is known, the new capacity expression can be obtained by approximating this transmit signal as a signal that is being transmitted through many small channels in frequency with small bandwidth. Even though the overall PSD of the signal is not flat, the PSD of each signal sent through each of the small channels can be approximated to be flat. Now by using the fundamental theorem of integral calculus, the total capacity for a given average SNR can be derived and is given by 
\begin{equation}
    C_{non-flat} = \int_{-\infty}^{\infty} \log_2\left(1+\text{SNR } W |H(f)|^2\right) df \text{ bits/s}
    \label{eq:c_non_flat}
\end{equation}
where, $H(f)$ is the normalized CTFT of the modulating pulse $h_{TX}(t)$.

It is shown in Appendix \ref{sec:append_cap_compare} that for any pulse $h_{TX}(t)$ that satisfies the Nyquist ISI criterion, $C_{non-flat} \geq C_{flat}$. This shows that a higher transmission rate is achievable in practical systems that do not use sinc pulses for modulation. This higher rate cannot be achieved by traditional signalling schemes where the symbols are spaced by period $T$ specified by the Nyquist ISI criterion. The following sections discuss how spacing the symbols closer than the period $T$ can allow for higher transmission rates.

\section{Capacity of Discrete-Time Systems}
\label{sec:dt_cap}
In the previous section, the capacity expression was given taking the received received signal $r(t)$ as continuous.
\begin{equation}
r(t) = \sum_m s[m]h_{TX}(t-mT) + \eta(t)
\end{equation}
where, $s[n]$ are the transmit symbols that have unit average power and are i.i.d. over $n$, $h_{TX}(t)$ is the modulating pulse used at the transmitter, $\eta(t)$ is the additive white Gaussian noise, and $T$ is the time period with which the symbols are spaced.

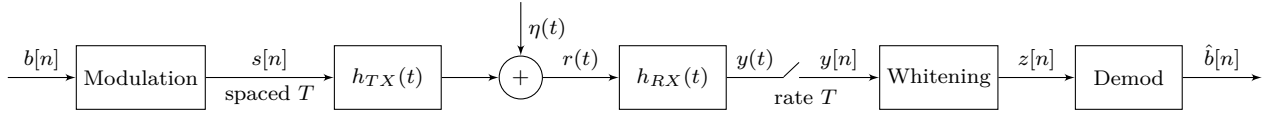
\begin{figure}[h]
\hspace{-1cm}
{\scriptsize
\begin{tikzpicture}[auto, >=latex']
\draw
    node [input] (input1) {} 
    node [block, right of=input1, node distance=1.75cm] (mod) {Modulation}
    node [block, right of=mod, node distance=3.25cm] (tx_pulse) {$h_{TX}(t)$}
    node [sum, right of=tx_pulse, node distance=1.75cm] (add_noise) {+}
    	node at (6.75, 1) [input] (input2) {}
    node [block, right of=add_noise, node distance=2cm] (matched_filter) {$h_{RX}(t)$}
    node [input, right of=matched_filter] (sampler) {}
    node [block, right of=sampler, node distance=2.5cm] (white) {Whitening}
    node [block, right of=white, node distance=2.5cm] (demod) {Demod}
	node [output, right of=demod, node distance=1.75cm] (output1){}
	;
	
	\sampler{0.75cm}{matched_filter}{sampler}
	\node at (10.5,-0.25) {rate $T$};
	
	\draw[->](input1) -- node {$b[n]$}(mod);
	\draw[->] (mod) -- node {$s[n]$} node [below] {spaced $T$} (tx_pulse);
	\draw[->](tx_pulse) -- node {}(add_noise);
	\draw[->](add_noise) -- node {$r(t)$}(matched_filter);
		\draw[->](input2) -- node {$\eta(t)$}(add_noise);
	\draw[-] (matched_filter) -- node {$y(t)$}(sampler);
	\draw[->](sampler) -- node {$y[n]$}(white);
 	\draw[->](white) -- node {$z[n]$} (demod);
	\draw[->](demod) -- node {$\hat{b}[n]$} (output1);
\end{tikzpicture}}
\caption{Block diagram of a modern digital communication system.}
\label{fig:discrete_time_block_diagram}
\end{figure}

However in a practical system, as shown in the block diagram in Fig. \ref{fig:discrete_time_block_diagram}, the received signal is sampled before demodulating the data.
Taking $h_{RX}(t)$ as the matched filter at the receiver and $\eta'(t)$ as the noise coloured due to the matched filter at receiver, the sampled signal at the receiver $y[n]$ i.e. the input to the whitening filter is given by
\begin{equation}
y[n] = y(nT) = \sum_m \left(h_{TX}*h_{RX}\right)(mT) s[n-m] + \eta'(nT)
\end{equation}
where, $y(t)$ is the output of matched filter at the receiver (see Fig. \ref{fig:discrete_time_block_diagram}) and $*$ represents convolution.

Assuming that the whitening filter used in this system is ideal, the output of the whitening filter is
\begin{equation}
z[n] = \sum_m h_{eff,T}[m] s[n-m] + w[n]
\label{eq:overall_discrete_time_system}
\end{equation}
where, $z[n]$ is the output of the whitening filter, $w[n]$ is the whitened noise, and $h_{eff,T}[n]$ is the effective response such that $h_{eff,T}[n]*h_{eff,T}[-n] = (h_{RX}*h_{TX})(nT)$. This response $h_{eff,T}[n]$ can be interpreted as the overall ISI introduced by the system. Here, the subscript $T$ signifies that the discrete sequences are obtained by sampling at rate $T$.

The capacity $C_{DT}$ of a discrete time system that has an input output relationship of (\ref{eq:overall_discrete_time_system}) and an overall filter response $h_{eff,T}[n]$ is given in \citep{info_rate} as
\begin{equation}
 C_{DT} = \frac{1}{2\pi}\int_{0}^{2\pi}\log_2\left(1+ \text{ SNR } \left|H_{eff,T}(e^{j\omega})\right|^2 \right) d\omega
\end{equation}
where, $H_{eff,T}(e^{j\omega})$ is the DTFT of $h_{eff,T}[n]$.

Denote $h_{eff,T}[n]*h_{eff,T}[-n]$ by $h_T[n]$ as 
\begin{equation}
h_T[n] = h_{eff,T}[n]*h_{eff,T}[-n] = (h_{RX}*h_{TX})(nT).
\end{equation}
Since $h_{eff,T}[n]*h_{eff,T}[-n] = h_T[n]$, 
\begin{equation}
\left|H_{eff,T}(e^{j\omega})\right|^2 = H_T(e^{j\omega})
\end{equation}
where,  $H_T(e^{j\omega})$ is the DTFT of the discrete sequence $h_T[n] = (h_{RX}*h_{TX})(nT)$. The capacity expression then simplifies to
\begin{equation}
\label{eq:c_dt}
     C_{DT} = \frac{1}{2\pi}\int_{0}^{2\pi}\log_2\left(1+ \text{ SNR } H_T(e^{j\omega}) \right) d\omega
\end{equation}
where, $H_T(e^{j\omega})$ is the DTFT of the discrete sequence $h_T[n] = (h_{RX}*h_{TX})(nT)$. 

Note that if the pulses used in the discrete time system are square root Nyquist, then 
\begin{equation}
h_T[n] = (h_{RX}*h_{TX})(nT) = \delta[n]
\end{equation}
because, the pulse $(h_{RX}*h_{TX})(t)$ satisfies Nyquist ISI criterion w.r.t. period $T$. Hence, the DTFT $H_T(e^{j\omega})$ of $h_T[n]$ is flat.
\begin{equation}
H_T(e^{j\omega}) = 1, \quad \forall \omega
\end{equation}
Plugging this in (\ref{eq:c_dt}),
\begin{equation}
C_{DT} = C_{flat}.
\end{equation}

This implies that, when the symbols are spaced at rate $T$, the capacity expression $C_{DT}$ is same as the standard capacity expression $C_{flat}$.
However, it was shown in the previous section that a higher capacity $C_{non-flat}$ is achievable in continuous-time systems when non-sinc pulses are used. To design a discrete-time system that has this higher capacity $C_{non-flat}$, the symbols need to be transmitted at a rate faster than the Nyquist (FTN) rate $T$. When the symbols are spaced at a smaller rate, the DFT $H_T(e^{j\omega}) $ is not flat any more, and hence $C_{DT}$ is not same as $C_{flat}$. The capacity of such an FTN system is derived in the following section. The next section also shows the comparison of capacity of FTN systems against $C_{flat}$ and $C_{non-flat}$.

\section{Capacity of FTN Systems}
\label{sec:ftn_cap}
In the previous section, the capacity expression for a discrete time system in which the symbols are spaced by the period $T$ was given. For a modulating pulse that was Nyquist with respect to this period $T$, it was seen that the capacity would be same as $C_{flat}$. In this section, the capacity of an FTN system is derived and shown how this capacity approaches the higher theoretical capacity $C_{non-flat}$ as the time acceleration factor $\tau$ approaches a particular value.

The block diagram of a single carrier FTN system is given in Fig. \ref{fig:ftn_sc_bd_cap}. 
\begin{figure}[h]
\hspace{-1cm}
{\scriptsize
\begin{tikzpicture}[auto, >=latex']
\draw
    node [input] (input1) {} 
    node [block, right of=input1, node distance=1.75cm] (mod) {Modulation}
    node [block, right of=mod, node distance=3.25cm] (tx_pulse) {$h_{TX}(t)$}
    node [sum, right of=tx_pulse, node distance=1.75cm] (add_noise) {+}
        node at (6.75, 1) [input] (input2) {}
    node [block, right of=add_noise, node distance=2cm] (matched_filter) {$h_{RX}(t)$}
    node [input, right of=matched_filter] (sampler) {}
    node [block, right of=sampler, node distance=2.5cm] (white) {Whitening}
    node [block, right of=white, node distance=2.5cm, text width = 1.5cm, align=center] (demod) {Equalize + Demod}
    node [output, right of=demod, node distance=1.75cm] (output1){}
    ;
    
    \sampler{0.75cm}{matched_filter}{sampler}
    \node at (10.5,-0.25) {rate $\tau T$};
    
    \draw[->](input1) -- node {$b[n]$}(mod);
    \draw[->] (mod) -- node {$s[n]$} node [below] {spaced $\tau T$} (tx_pulse);
    \draw[->](tx_pulse) -- node {}(add_noise);
    \draw[->](add_noise) -- node {$r(t)$}(matched_filter);
        \draw[->](input2) -- node {$\eta(t)$}(add_noise);
    \draw[-] (matched_filter) -- node {$y(t)$}(sampler);
    \draw[->](sampler) -- node {$y[n]$}(white);
    \draw[->](white) -- node {$z[n]$} (demod);
    \draw[->](demod) -- node {$\hat{b}[n]$} (output1);
\end{tikzpicture}}
\caption{Block diagram of a single carrier FTN communication system.}
\label{fig:ftn_sc_bd_cap}
\end{figure}
Here, the modulating pulse is Nyquist w.r.t. to the period $T$ but the symbols are spaced by  $\tau T$ $(< T)$. Accordingly, at the receiver, the sampling is done at a faster rate of $\tau T$. The sampled signal $y[n]$ at the receiver, i.e. the input to the whitening filter, is given by
\begin{equation}
y[n] = y(n\tau T) = \sum_m \left(h_{TX}*h_{RX}\right)(m\tau T) s[n-m] + \eta'[n]
\end{equation}
where, $y(t)$ is the output of matched filter at the receiver (see Fig. \ref{fig:ftn_sc_bd_cap}), $y[n]$ is the sampled signal at the receiver that is sampled at a rate of $\tau T$, $h_{RX}(t)$ is the matched filter at the receiver, $*$ represents convolution, $\eta'[n] = \eta'(n\tau T)$, and $\eta'(t)$ is the coloured noise coloured due to the matched filer at receiver.

Assuming that the whitening filter used in this system is ideal, the output of the whitening filter is
\begin{equation}
z[n] = \sum_m h_{eff}[m] s[n-m] + w[n]
\end{equation}
where, $z[n]$ is the output of the whitening filter, $w[n]$ is the whitened noise, and $h_{eff}[n]$ is the effective response such that $h_{eff}[n]*h_{eff}[-n] = (h_{RX}*h_{TX})(n\tau T)$. Note here that, unlike the previous section, the subscript $T$ is dropped. As the sampling is done at rate $\tau T$, the subscript should have been $\tau T$, but this subscript is dropped for better readability.

 Using the capacity expression (\ref{eq:c_dt}), the capacity of this FTN system is given by
\begin{equation}
C_{FTN} = \frac{1}{2\pi}\int_{0}^{2\pi}\log_2\left(1+ \tau \text{ SNR } H(e^{j\omega}) \right) d\omega
\label{eq:c_ftn}
\end{equation} 
where, $H(e^{j\omega})$ is the DTFT of the discrete sequence $h[n] = (h_{RX}*h_{TX})(n\tau T)$. As the pulse $(h_{RX}*h_{TX})(t)$ is Nyquist w.r.t. $T$, the DTFT $H(e^{j\omega})$ is not flat. Note here that the SNR is scaled by a factor of $\tau$ to account for the increase in noise level due to denser sampling.  

In the following subsections, the variation of the FTN capacity $C_{FTN}$ as a function of $\tau$ is examined. First, the analysis is done assuming the modulating pulse to be SRRC. Later, the analysis is done assuming the modulating pulse to be  rectangular.

\subsection{Capacity of FTN systems that use SRRC modulating pulse}
To get the capacity of an FTN system that uses SRRC pulse for modulation, plug in $h_{RX}(t) = h_{TX}(t) = h_{SRRC}(t)$ where, $h_{SRRC}(t)$ is the square-root raised cosine pulse. The discrete sequence $h[n]$ in (\ref{eq:c_ftn}) can be simplified as
\begin{equation}
h[n] = (h_{RX}*h_{TX})(n\tau T) = (h_{SRRC}*h_{SRRC})(n\tau T) = h_{RC}(n\tau T)
\end{equation}
where, $h_{RC}(t)$ is the raised cosine pulse. Hence, the DTFT $H(e^{j\omega})$ of $h[n]$ present in the capacity expression can be replaced with the DTFT $H_{RC}(e^{j\omega})$ of the RC pulse. The capacity expression is
\begin{equation}
C_{FTN,SRRC} = \frac{1}{2\pi}\int_{0}^{2\pi}\log_2\left(1+ \tau\text{ SNR } H_{RC}(e^{j\omega}) \right) d\omega.
\end{equation}

The DTFT $H_{RC}(e^{j\omega})$ of the RC pulse can be expressed in terms of the CTFT $\mathcal{H}_{RC}(f)$ of the RC pulse as
\begin{equation}
{H_{RC}(e^{j\omega})} = \frac{1}{\tau T}\sum_{k=-\infty}^\infty \mathcal{H}_{RC}\left(\frac{\omega-2\pi k}{2\pi \tau T}\right).
\end{equation}

By plugging in the above expression for the DTFT and simplifying, the capacity can be expressed in terms of the CTFT as
\begin{equation}
\label{eq:c_ftn_srrc}
C_{FTN,SRRC} = \int_{0}^{1/2\tau T}\log_2\left(1+ \text{ SNR } \sum_{k=-\infty}^\infty \frac{1}{T} \mathcal{H}_{RC}\left(f-\frac{k}{\tau T}\right) \right) df \text{ bits/s}.
\end{equation}

\subsubsection{Variation of capacity $C_{FTN,SRRC}$ with value of $\tau$}

In order to understand the dependency of the FTN capacity on $\tau$, two things are to be noted in the capacity expression of (\ref{eq:c_ftn_srrc}).
\begin{enumerate}
\item The integration limits are from $0$ to $1/2\tau T$.
\item The integrand is a function of multiple copies of the CTFT $\mathcal{H}_{RC}(f)$ shifted by integral multiples of $1/\tau T$.
\end{enumerate}

\begin{figure}[h]
    \centering
    \begin{subfigure}{0.45\textwidth}
        \centering
        \includegraphics[width=\textwidth]{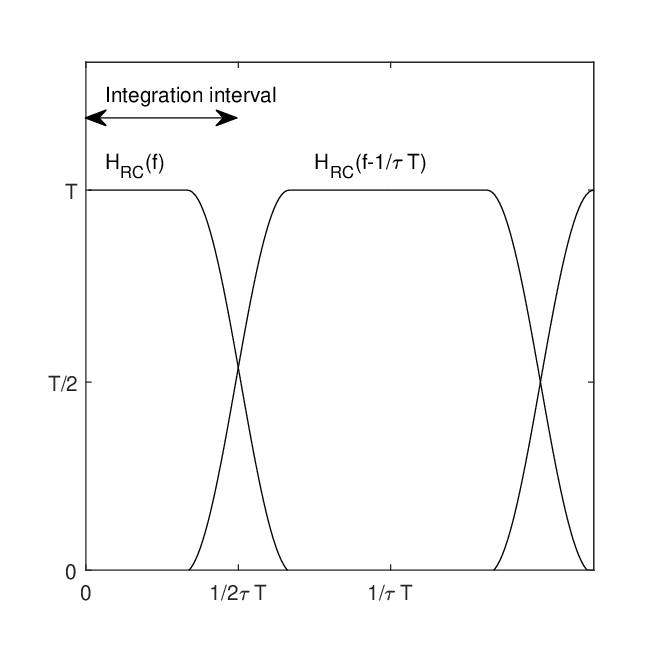}
        \caption{$\tau=1$}
    \end{subfigure}
    ~
    \begin{subfigure}{0.45\textwidth}
        \centering
        \includegraphics[width=\textwidth]{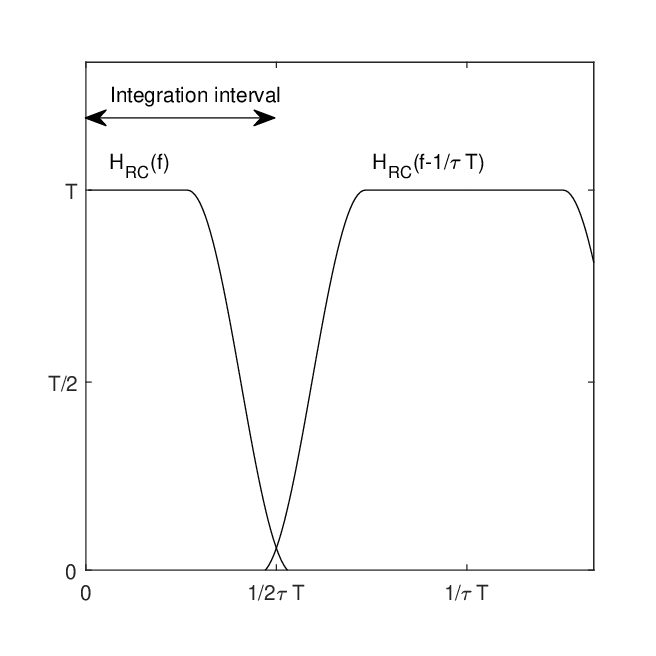}
        \caption{$\tau=0.8$}
    \end{subfigure}
    
    \begin{subfigure}{0.45\textwidth}
        \centering
        \includegraphics[width=\textwidth]{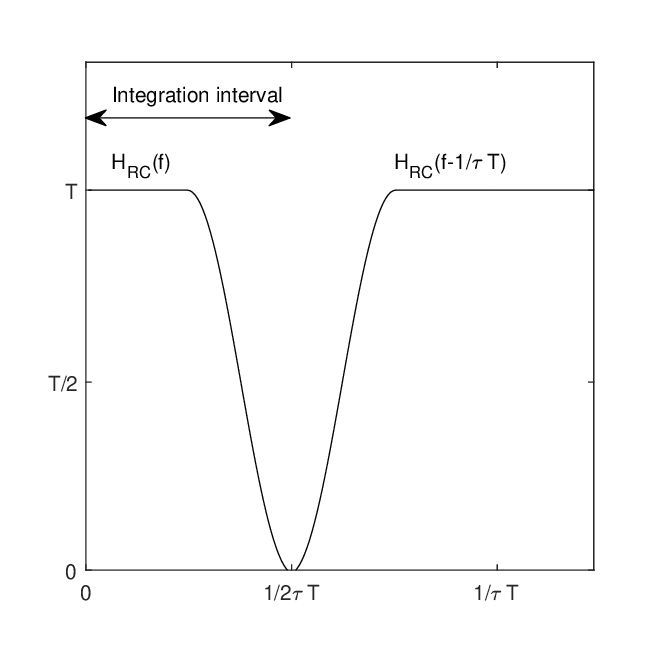}
        \caption{ $\tau=1/(1+\alpha)=0.74$}
    \end{subfigure}
    ~
    \begin{subfigure}{0.45\textwidth}
        \centering
        \includegraphics[width=\textwidth]{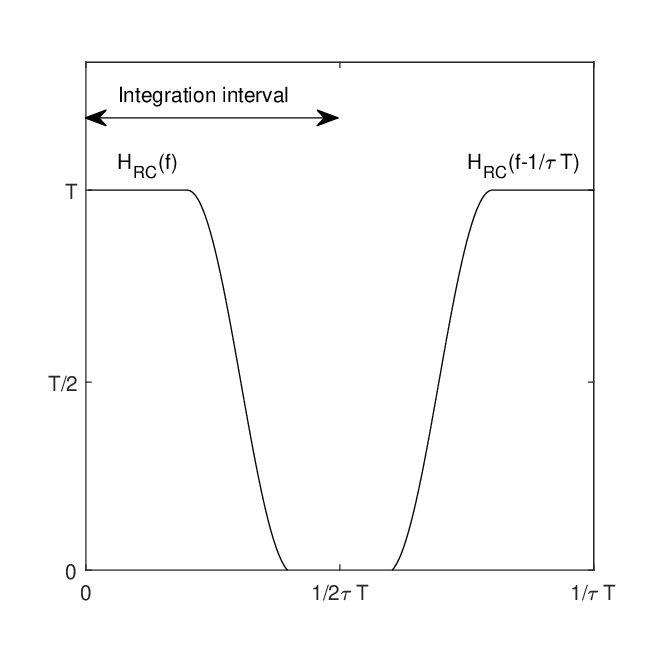}
        \caption{$\tau=0.6$}
    \end{subfigure}
    
    \caption{The copies of the CTFT $\mathcal{H}_{RC}(f)$ of the RC pulse with roll-off $\alpha=0.3$ separated by $1/\tau T$ and the interval of integration for different values of $\tau$.}
    \label{fig:integration_curves}
\end{figure}

Fig. \ref{fig:integration_curves} shows the shifted copies of the CTFT $\mathcal{H}_{RC}(f)$ of an RC pulse with roll-off $\alpha = 0.3$ for different values of $\tau$. It also marks the limits of integration for the capacity calculation as in (\ref{eq:c_ftn_srrc}).

From Fig. \ref{fig:integration_curves}, it can be seen that as the value of $\tau$ decreases, the interval of integration becomes wider and hence, as long as $\tau > 1/(1+\alpha)$, the capacity of the system increases. This is shown formally in Appendix \ref{sec:append_cap_ftn_srrc} by calculating the derivative $dC_{FTN,SRRC}/d\tau$ and showing that the derivative is negative for all $\tau>1/(1+\alpha)$. Note that this increase in capacity with decreasing values of $\tau$ is valid only when $\tau > 1/(1+\alpha)$. The case of $\tau <  1/(1+\alpha)$ is discussed in the following paragraph.

As the value of $\tau$ decreases, the separation $1/\tau T$  between every adjacent copies increases. When the value of $\tau$ is $\tau = 1/(1+\alpha)$, the separation between every adjacent copies is equal to $(1+\alpha)/T$. Since each copy is bandlimited to $(1+\alpha)/2T$, the copies are completely separated from each other in frequency at this value of $\tau = 1/(1+\alpha)$. Further, for values of $\tau < 1/(1+\alpha)$, the separation between every adjacent copies is larger than $(1+\alpha)/T$. Hence, the copies are completely separated from each other for all values of $\tau < 1/(1+\alpha)$.  

From the above explanation and from Fig. \ref{fig:integration_curves}, the following observation can be made. As the value of $\tau$ decreases below  $1/(1+\alpha)$, even though the interval of integration becomes wider, the adjacent copies are completely separate from each other and hence the value of integrand in the region of widened interval of integration is zero. Hence, the capacity does not increase with $\tau$ for $\tau<1/(1+\alpha)$. This is formally shown in Appendix \ref{sec:append_cap_ftn_srrc} by calculating the derivative $dC_{FTN,SRRC}/d\tau$ and showing that it is zero for all $\tau<1/(1+\alpha)$.

This limit on $\tau$, below which reducing the value of $\tau$ does not have any effect on improving the capacity, reinforces the limits derived in Chapter \ref{chap:isi_perspective}. This limit $\tau = 1/(1+\alpha)$ is same as the limit derived in Chapter \ref{chap:isi_perspective}. In Chapter \ref{chap:isi_perspective}, it was derived that,  if  $\tau < 1/(1+\alpha)$, then the ISI introduced by FTN cannot be mitigated completely and hence it was unnecessary to reduce the value of $\tau$ below this limit.

\subsubsection{Capacity $C_{FTN,SRRC}$ of mild FTN systems (values of $\tau \approx 1$)}
To see what is the lower limit on the capacity of FTN systems using SRRC modulating pulse, the capacity of very mild FTN system needs to be calculated. To do this, the capacity expression in (\ref{eq:c_ftn_srrc}) is calculated at $\tau = 1$.
\begin{equation}
\begin{split}
C_{FTN,SRRC} & = \int_{0}^{1/2\tau T}\log_2\left(1+ \text{ SNR } \sum_{k=-\infty}^\infty \frac{1}{T} \mathcal{H}_{RC}\left(f-\frac{k}{\tau T}\right) \right) df \\
& =  \int_{0}^{1/2 T}\log_2\left(1+ \text{ SNR } \sum_{k=-\infty}^\infty \frac{1}{T} \mathcal{H}_{RC}\left(f-\frac{k}{ T}\right) \right) df \text{ bits/s.}
\end{split}
\end{equation}

Since, RC pulse is Nyquist w.r.t. $\tau$, from the Nyquist zero-ISI criterion,
\begin{equation}
\frac{1}{T}\sum_{k=-\infty}^\infty \mathcal{H}_{RC}\left(f-\frac{k}{ T}\right) = 1, \quad \forall f. 
\end{equation}

Hence, the capacity expression simplifies to
\begin{equation}
C_{FTN,SRRC} =  \int_{0}^{1/2 T}\log_2\left(1+ \text{SNR}\right) df = W\log_2(1+\text{SNR}) = C_{flat}
\end{equation}
for $\tau = 1$.

As it is already shown that the capacity of FTN system increases with decreasing $\tau$, it can be concluded that,
\begin{equation}
\label{eq:ftn_cap_lower_lim}
C_{FTN,SRRC} \geq C_{flat}, \quad \forall \tau \leq 1. 
\end{equation}

\subsubsection{Capacity $C_{FTN,SRRC}$ of severe FTN systems (small values of $\tau$)}

To see what is the maximum capacity that an FTN system can reach using SRRC modulating pulse, the capacity expression in (\ref{eq:c_ftn_srrc}) is calculated in the limit as $\tau$ approaches zero. As $\tau$ approaches zero, the separation between the first copy of the CTFT and rest of the copies goes to infinity. Hence, the capacity value in the limit is given by
\begin{equation}
\lim_{\tau \rightarrow 0}C_{FTN,SRRC} =  \int_{0}^{\infty}\log_2\left(1+ \text{ SNR } \frac{1}{T} \mathcal{H}_{RC}(f) \right) df \text{ bits/s}.
\end{equation}

It can be seen that this capacity is of the form $C_{non-flat}$ in (\ref{eq:c_non_flat}). Combining this result with (\ref{eq:ftn_cap_lower_lim}), it can be concluded that $C_{non-flat} \geq C_{FTN,SRRC} \geq C_{flat}$.

\subsubsection{Plots of capacity $C_{FTN,SRRC}$ versus SNR}

\begin{figure*}[htbp]
    \centering
    \begin{subfigure}[t]{0.48\textwidth}
        \centering
        \includegraphics[width=\textwidth]{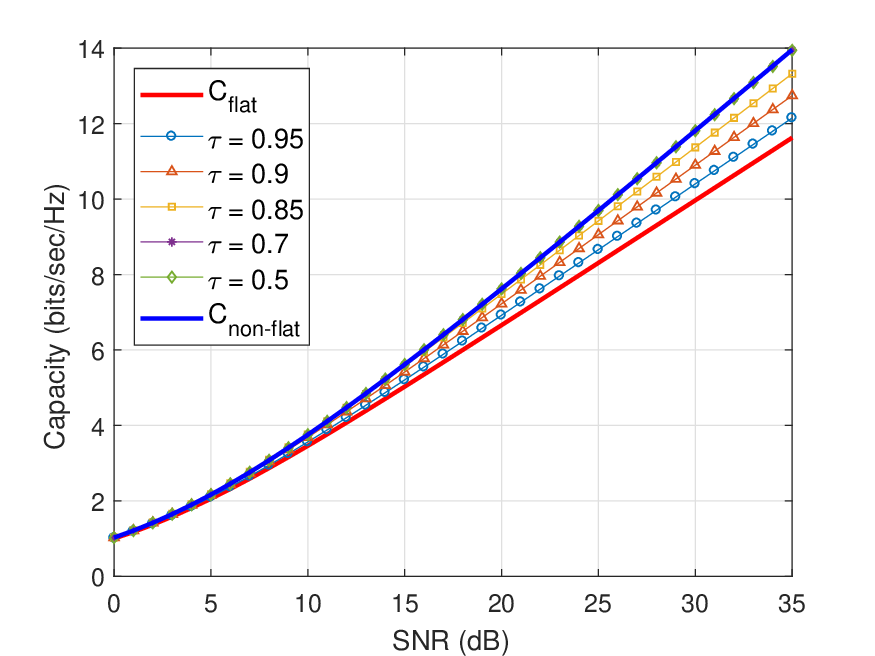}
        \caption{For SNR range 0 to 35 dB.}
    \end{subfigure}
    ~
    \begin{subfigure}[t]{0.48\textwidth}
        \centering
        \includegraphics[width=\textwidth]{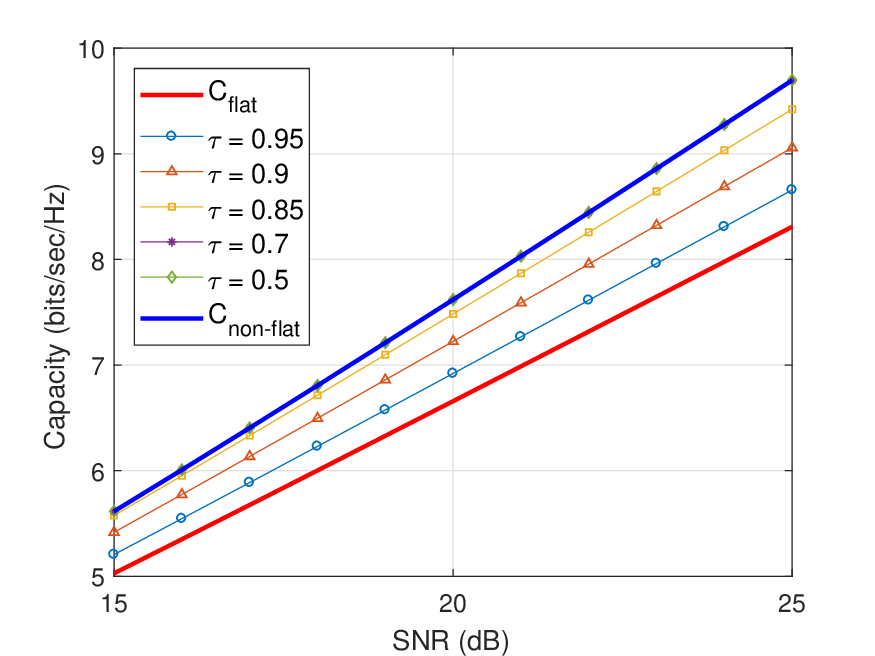} 
        \caption{For SNR range 15 to 25 dB (zoomed).}
    \end{subfigure}
    \caption{Capacity curves of FTN system with SRRC pulse of roll-off $\alpha = 0.3$.}
    \label{fig:capacity_plot_srrc}
\end{figure*}

Finally, in order to verify the above explained behaviour, the capacity expression in (\ref{eq:c_ftn_srrc}) was plotted against SNR by solving the integration numerically. The plot is shown in Fig. \ref{fig:capacity_plot_srrc}.
It can be seen that, for a particular value of SNR, the capacity increases with decreasing $\tau$ till $\tau > 1/(1+\alpha)$. And for values of $\tau <  1/(1+\alpha)$, the capacity saturates to $C_{non-flat}$ and remains constant with $\tau$.

\subsection{Capacity of FTN systems that use rectangular modulating pulse}
In OFDM digital communication systems, rectangular pulses are often used for modulation \citep{Zhao2017PulseSD}. So, analysing the FTN capacity in systems  that use rectangular modulating pulses is necessary. The capacity expression (\ref{eq:c_ftn_srrc}) used for the case of SRRC pulse can be used in this case with slight modifications. When rectangular pulse is used instead of SRRC pulse, the overall response $h[n]$ in the FTN capacity expression (\ref{eq:c_ftn}) is given by 
\begin{equation}
h[n] = (h_{RX}*h_{TX})(n\tau T) = (h_{rect}*h_{rect})(n\tau T) = h_{tri}(n\tau T)
\end{equation}
where, $h_{rect}(t)$ is the rectangular pulse and $h_{tri}(t)$ is the triangular pulse that is Nyquist w.r.t. $T$.

Now, since the CTFT of a triangular pulse is given by sinc squared, the final capacity expression can be derived from (\ref{eq:c_ftn_srrc}) as
\begin{equation}
\label{eq:c_ftn_rect}
C_{FTN,rect} = \int_{0}^{1/2\tau T}\log_2\left(1+ \text{ SNR } \sum_{k=-\infty}^\infty \frac{1}{T} \mathcal{H}_{tri}\left(f-\frac{k}{\tau T}\right) \right) df \text{ bits/s}
\end{equation}
where,
\begin{equation}
\label{eq:triangular_ft}
    \mathcal{H}_{tri}(f) = \left(\frac{\sin (\pi f T)}{\pi f T}\right)^2 .
\end{equation}

Now, this capacity expression can be analysed in a similar way as done in the case of SRRC pulse. As the value of $\tau$ decreases, the interval of integration increases. In case of SRRC pulse, there was a limit on $\tau$ below which the adjacent copies of CTFT were separated completely and hence, did not contribute to an increase in capacity. However, the CTFT of the rectangular pulse is sinc in frequency domain, which is not frequency limited. Hence, as the value of $\tau$ decreases, the adjacent copies never get completely separated. Hence, unlike the case of SRRC pulse, the value of capacity in the case of rectangular pulse does not saturate at a particular value of $\tau$.

\begin{figure*}[htbp]
    \centering
    \begin{subfigure}[t]{0.48\textwidth}
        \centering
        \includegraphics[width=\textwidth]{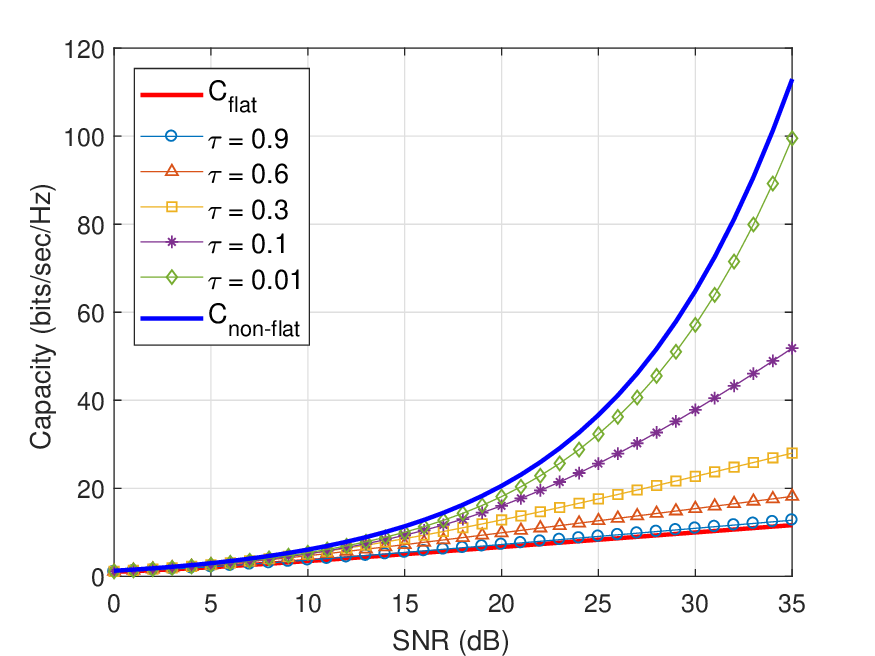}
        \caption{For SNR range 0 to 35 dB.}
    \end{subfigure}
    ~
    \begin{subfigure}[t]{0.48\textwidth}
        \centering
        \includegraphics[width=\textwidth]{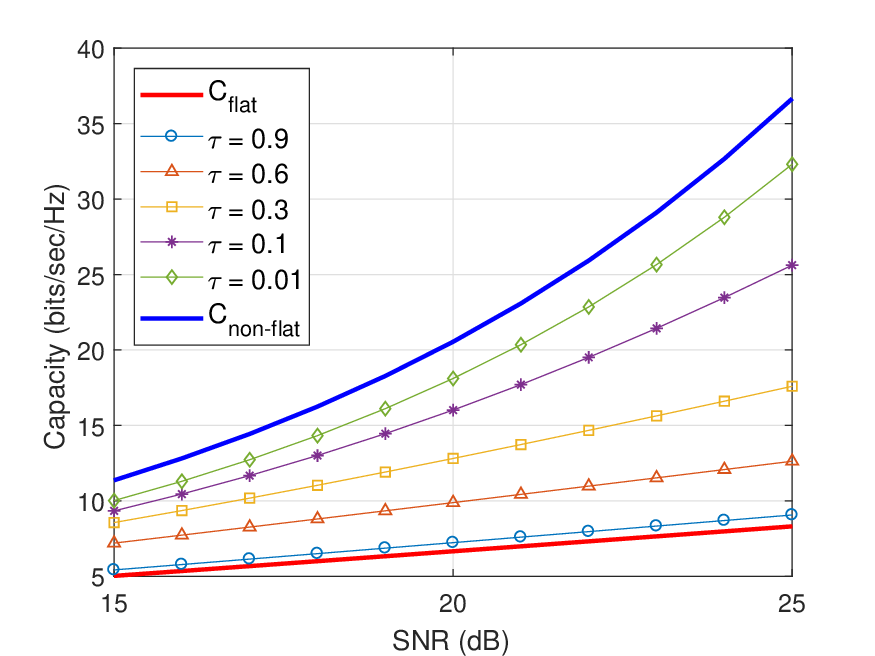}
        \caption{For SNR range 15 to 25 dB (zoomed).}
    \end{subfigure}
    \caption{Capacity curves of the FTN system with rectangular pulse.}
    \label{fig:cap_rect}
\end{figure*}

The capacity expression in (\ref{eq:c_ftn_rect}) plotted against SNR is shown in Fig.  \ref{fig:cap_rect}.
It can be seen that the capacity values keep on increasing as the value of $\tau$ decreases and approaches the saturation value of $C_{non-flat}$ for very low values of $\tau$.

Now that the capacity expressions for FTN systems are derived and shown to be higher than Nyquist systems, the next chapter discusses few techniques that can be used to reach these capacity rates. These techniques take advantage of the OFDM structure and reduce the effect of ISI and increase the FTN throughput.


\chapter{POWER ALLOCATION AND ADAPTIVE LOADING}
\label{chap:waterfilling_loading}
In Chapter \ref{chap:capacity_perspective}, it was shown that the FTN signalling scheme can reach a higher capacity than the traditional signalling scheme. However, in order to reach this higher capacity, various techniques like error control coding, water-filling, and loading are required. This chapter discusses few of these techniques that can be employed to increase the transmission rates of the FTN systems.

In Chapter \ref{chap:isi_perspective}, the conditions on modulating pulse were derived for complete ISI mitigation. It was shown that the ISI can be inverted completely when the DTFT of this multi-tap channel is non-zero for all frequency. The condition was also derived considering a multi-carrier OFDM FTN system of the following form.
\begin{equation}
    y[i] = H[i]s[i] + \eta'[i], \quad i={0,\dots,N-1} 
\end{equation}
where $H[i]$ is the $i$-th coefficient of N-point DFT of $(h_{TX}*h_{RX})(t)|_{\tau T}$ and $\eta'[i]$ is the coloured noise. It was shown that in order to estimate $s[i]$ from $y[i]$, none of the $H[i]$ should be equal to zero i.e. none of the DFT coefficients of $(h_{TX}*h_{RX})(t)|_{\tau T}$ should be zero. The condition was derived on the values of time acceleration factor $\tau$ and the roll-off $\alpha$ as $(1+\alpha)\tau>1$.

If this condition is not satisfied, then there exists some frequencies $\omega$ where $H_{RC}(e^{j\omega})$ goes to zero (see Fig. \ref{fig:SC_channel_noop} for example). The sub-carriers corresponding to those frequencies have $H[i]=0$. Thus, those sub-carriers cannot be used to transmit symbols. The natural idea that arises now is the following. By not transmitting any symbols in those sub-carriers, the transmit power that was originally allocated to those sub-carriers can be reallocated to other non-zero sub-carriers. This naturally leads to the optimal power allocation scheme that is obtained from the water-filling algorithm.

\section{Power Allocation in Multi-Carrier FTN systems with the Water-filling Algorithm} 
\label{sec:power_alloc}
The idea behind power allocation is to vary power assigned to each sub-channel relative to that sub-channel gain. It was seen in Chapter \ref{chap:isi_perspective}, Section \ref{sec:analysis_ofdm} that, in an FTN OFDM system with time acceleration factor $\tau$ and modulating pulses $h_{TX}(t)$ and $h_{RX}(t)$ at the transmitter and the receiver, the sub-channel gains are given by $H[i]$, $i={0,\dots,N-1}$, the N-point DFT coefficients of the sequence  $(h_{TX}*h_{RX})(t)|_{\tau T}$. Now, the power allocation scheme can be obtained by maximizing the capacity expression considering this sub-channel gain. The power allocation $P_i$ for each sub-carrier $i$ is given by
\begin{equation}
P_i = \argmax_{P_i\: :\:\sum_i P_i < \overline{P}} \;\; \sum_{i=0}^{N-1} W \log(1+ P_i H[i]^2 \text{ SNR}), \quad  i={0,\dots,N-1}
\end{equation} 
where, $\overline{P}$ is the average power and $\sum_i P_i < \overline{P}$ is the average power constraint. The optimal power allocation obtained by solving \citep{goldsmith_2005} this optimization problem is 
\begin{equation}
\frac{P_i}{\overline{P}} = \begin{cases} 
      1/\gamma_0 - 1/\gamma_i  & \text{ for } \gamma_i \geq \gamma_0 \\
      0 & \text{ for } \gamma_i < \gamma_0 
\end{cases}
\end{equation}
where, $\gamma_i$ is the SNR corresponding to each sub-channel given by $\gamma_i = H[i]^2 \text{ SNR}$ and $\gamma_0$ is a cut-off obtained from solving the constraint $\sum_i P_i = \overline{P}$.

The simulation results showing the variation in transmission rates with power allocation is shown in Chapter \ref{chap:simulation}, Section \ref{sec:wf_sim}.

The natural extension to the idea of power allocation is adaptive bit loading. In power allocation, each sub-carrier used a different transmit power depending on the SNR value of that particular sub-carrier. In adaptive loading, each sub-carrier uses a different modulation scheme depending on the SNR value of that particular sub-carrier. The following section discusses the details of bit adaptive loading and how it improves the transmission rates in FTN systems.

\section{Adaptive Bit Loading in Multi-Carrier FTN systems} 
\label{sec:loading}
In FTN systems with OFDM modulation, the frequency varying sub-channel gains were given by the DFT of the sampled version of the modulating pulse. In the previous section, the method of power allocation across different sub-carriers was discussed. The natural extension to the idea of power allocation is adaptive bit loading. In adaptive loading, each sub-carrier uses a different modulation scheme depending on the SNR value of that particular sub-carrier. It is well known that the higher modulation schemes (i.e. schemes with larger constellations) provide higher throughputs at high SNR regions. The idea behind adaptive loading is to use higher modulations in sub-carriers that have higher SNR. The following steps can be followed to implement such an adaptive loading technique.

The packet throughput values for a range of SNR is obtained from simple AWGN channel simulations. At every SNR $\gamma$, packet throughput values of different modulations are obtained. Now, to decide the modulation scheme to be used in the $i$-th sub-carrier in an OFDM FTN system, the SNR of this sub-channel is calculated as $\gamma_i = H[i]^2 \text{ SNR}$. For this value of SNR, the packet throughput values for different modulations obtained from the simulations are examined. Among these modulations, the modulation that has the highest throughput is selected. 

A more detailed explanation of the implementation is given in Chapter \ref{chap:simulation}, Section \ref{sec:loading_sim}. The simulation results showing the improvement in the transmission rates of FTN systems by using adaptive bit loading is also shown in Section \ref{sec:loading_sim}.
Finally, a simulation is performed for OFDM FTN systems employing both the power allocation and adaptive loading techniques discussed in this chapter. The simulation results comparing the transmission rates of these systems with different values of $\tau$ against Nyquist signalling is shown in Section \ref{sec:final_sim}.


\chapter{SIMULATION RESULTS}
\label{chap:simulation}
The power allocation and adaptive loading techniques discussed in the previous chapter were implemented in MATLAB. The results of these simulations are discussed in this chapter. All the simulations done below are for OFDM systems. The modulating pulse used in these simulations were SRRC with roll-off $\alpha = 0.3$.

\section{Throughput Simulations with Power Allocation}
\label{sec:wf_sim}
The power allocation scheme discussed in Chapter \ref{chap:waterfilling_loading}, Section \ref{sec:power_alloc} was implemented and simulations were performed for different modulation schemes. The pseudocode for the implementation is given in Appendix \ref{sec:append_waterfilling}.  Fig. \ref{fig:waterfilling} shows the results of simulation where the resulting packet throughput is plotted against SNR.
\begin{figure*}[h]
    \centering
    \begin{subfigure}{0.48\textwidth}
        \centering
        \includegraphics[width=\textwidth]{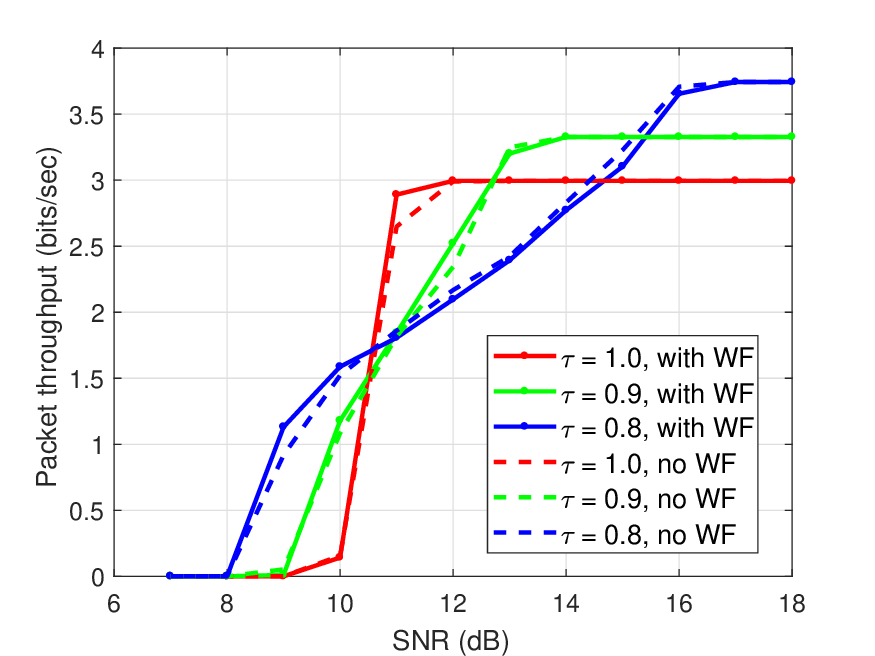}
        \caption{64 QAM}
    \end{subfigure}
    ~
    \begin{subfigure}{0.48\textwidth}
        \centering
        \includegraphics[width=\textwidth]{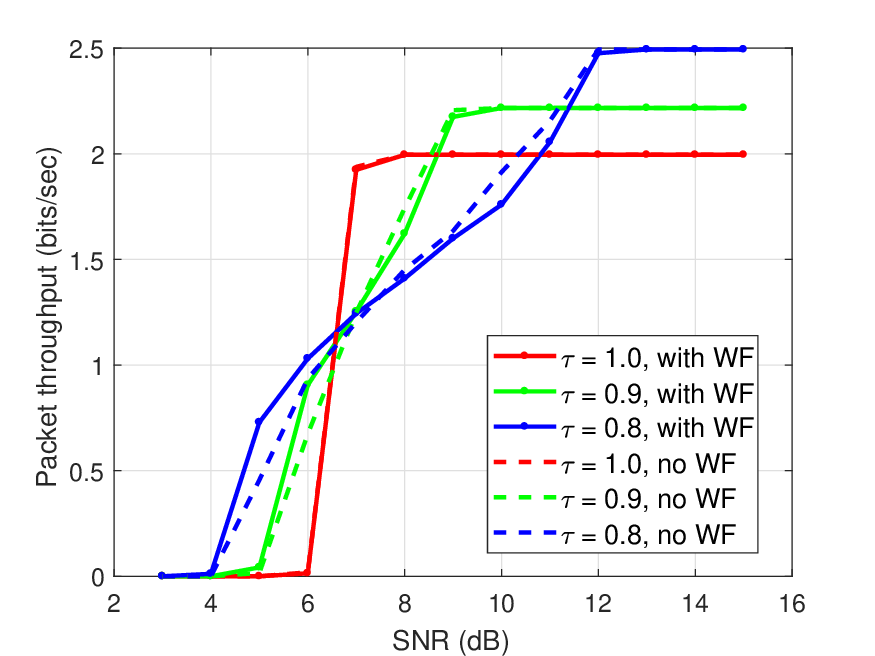}
        \caption{16 QAM}
    \end{subfigure}
    
    \begin{subfigure}{0.48\textwidth}
        \centering
        \includegraphics[width=\textwidth]{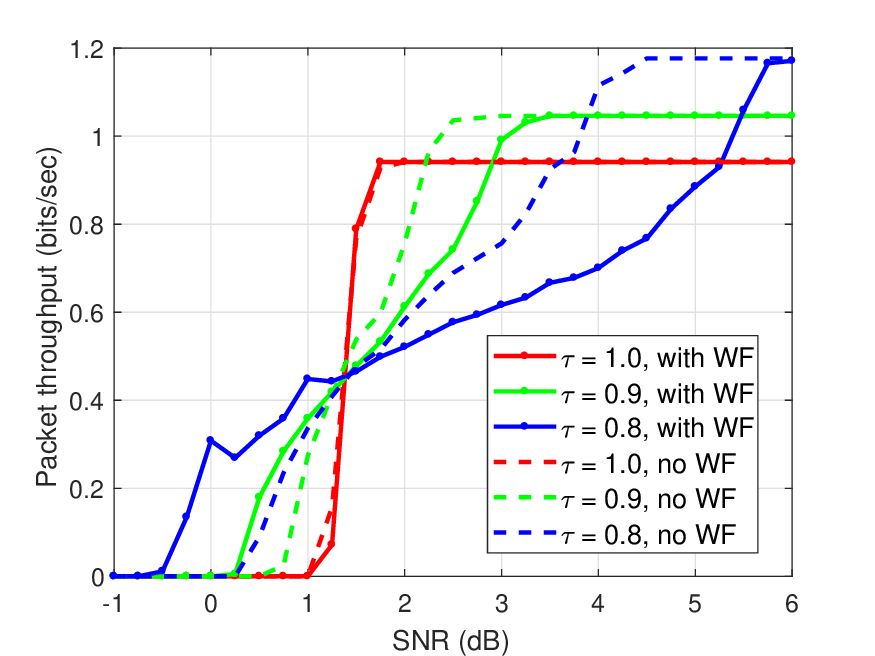}
        \caption{QPSK}
    \end{subfigure}
    ~
    \begin{subfigure}{0.48\textwidth}
        \centering
        \includegraphics[width=\textwidth]{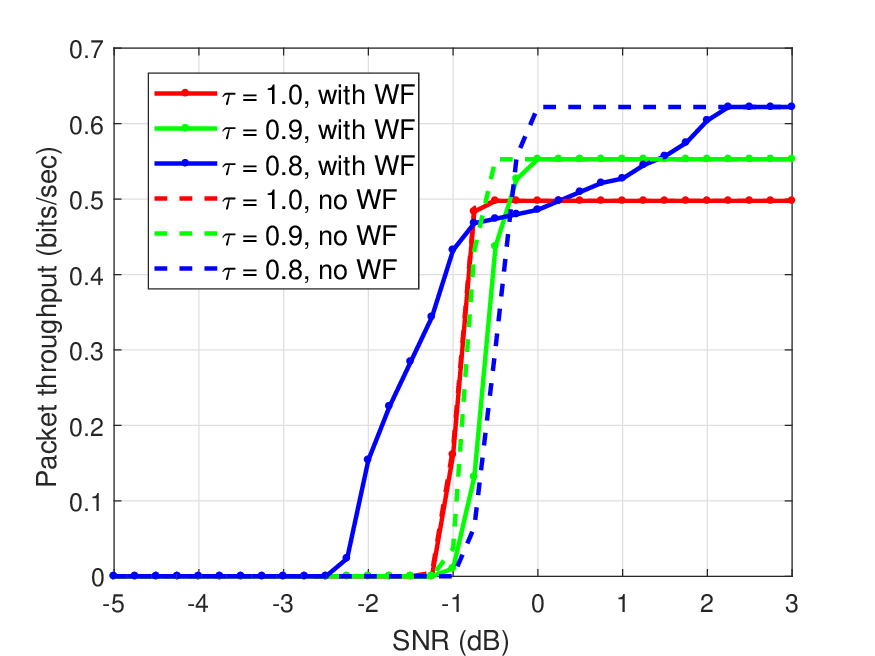}
        \caption{BPSK}
    \end{subfigure}
    \caption{Simulation results to show the variation of packet throughput for different modulations with and without employing the water-filling algorithm. In the plots, WF stands for water-filling.}
    \label{fig:waterfilling}
\end{figure*}
Simulations are done for different modulation schemes like 64QAM, 16QAM, QPSK, and BPSK. For each of these modulation schemes, simulations are performed with and without employing the water-filling algorithm. Further, for both the cases of with and without water-filling, the packet throughputs are shown for different values of $\tau = \{1.0, 0.9, 0.8\}$.

To understand the results in Fig. \ref{fig:waterfilling}, first, the throughput curves with no water-filling (dotted curves) are examined. For $\tau = 1$, the throughput curves start raising at a particular SNR, they raise rapidly and reach their saturation values. For $\tau < 1$, the throughput curves start raising at an earlier SNR, they raise less rapidly, and reach higher saturation values.

Now, when the water-filling algorithm is employed, the  transition of throughput from zero to saturation becomes more skewed. The throughput at lower SNR improves, but throughput at higher SNR worsens. In order to ensure that the throughput improves at higher SNR regions too, the adaptive loading technique is required. The simulations for the loading technique and simulations combining both the power allocation and adaptive loading are shown in the following sections.

\section{Throughput Simulations with Adaptive Loading}
\label{sec:loading_sim}
Adaptive bit loading technique was discussed in Chapter \ref{chap:waterfilling_loading}, Section \ref{sec:loading}. In adaptive loading, each sub-carrier uses a different modulation scheme depending on the SNR value of that particular sub-carrier. As explained in  Section \ref{sec:loading}, in order to decide the modulation scheme, simulations are performed on simple AWGN channels beforehand. These simulations can later be used to decide the type of modulation at each sub-carrier.

\begin{figure}[htbp]
    \centering
    \includegraphics[width = 0.75\textwidth]{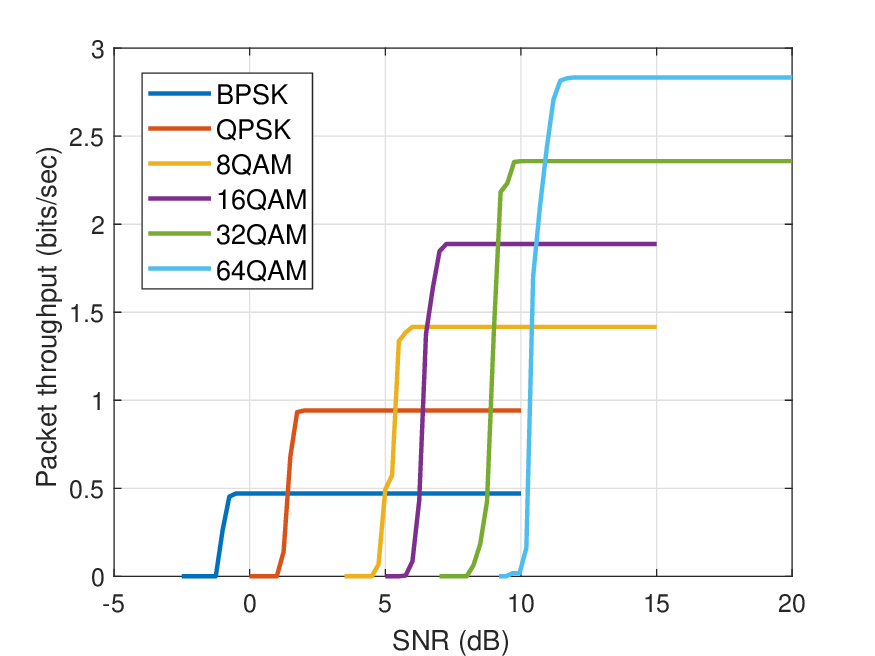}
    \caption{Simulation of different modulation schemes to be used as reference while performing adaptive loading. The plot shows packet throughput values against SNR for different modulation schemes.}
    \label{fig:basethroughput}
\end{figure}

The pseudocode of this simulation is given in Appendix \ref{sec:append_baseline}. Transmit bits are sent from the transmitter on a single tap channel. AWGN noise depending on the SNR is added to the transmit symbols and received at the receiver. Packet throughputs are computed and averaged over multiple iterations. The result of this simulation is shown in Fig. \ref{fig:basethroughput}. 
The simulation is done for various QAM modulation schemes. For each modulation scheme, the packet throughputs are plotted against SNR.

It can be seen from the figure that at lower SNR values, lower modulation schemes like BPSK provide higher throughput than other modulation schemes. This is because, at lower SNR region, higher modulation schemes, despite having higher rate, experience severe error rates. Where as, at higher SNR region, higher modulation schemes like 64QAM provide higher throughput as expected.

Now, the thresholds as given in Table \ref{tab:loading_thresholds} can be decided based on this simulation.
\begin{table}[h]
\begin{center}
\begin{tabular}{cc}
\hline
SNR range & Modulation scheme\rule{0pt}{2.6ex}\rule[-0.9ex]{0pt}{0pt} \\
\hline\hline
$< 1.5$ dB & BPSK\rule{0pt}{2.6ex} \\
$1.5$ dB to $5.5$ dB & QPSK \\
$5.5$ dB to $6.5$ dB & 8QAM \\
$6.5$ dB to $9.5$ dB & 16QAM \\
$9.5$ dB to $11.2$ dB & 32QAM \\
$> 11.2$ dB & 64QAM\rule[-0.9ex]{0pt}{0pt} \\
\hline
\end{tabular}
\end{center}
\caption{Modulation schemes that give the best packet throughput for a given SNR. These values are decided from the simulation results obtained in Fig. \ref{fig:basethroughput}.}
\label{tab:loading_thresholds}
\end{table}
For SNR values lesser than $1.5$ dB, it can be seen from Fig. \ref{fig:basethroughput} that BPSK has the highest throughput compared to other modulation schemes. Hence, BPSK is chosen for this SNR range. Similarly for SNR range of $1.5$ dB to $5.5$ dB, QPSK has the best throughput compared to other modulation schemes. Similarly, rest of the table is filled. Note that modulation schemes higher than 64QAM can be considered for higher SNR values. In this thesis, however, only modulation schemes till 64QAM are considered.

Now that the modulation scheme that best suits for each SNR value is known, adaptive bit loading can be performed for OFDM FTN systems. In an FTN OFDM system with time acceleration factor $\tau$ and modulating pulses $h_{TX}(t)$ and $h_{RX}(t)$ at the transmitter and the receiver, the sub-channel gains are given by $H[i]$, $i={0,\dots,N-1}$, the N-point DFT coefficients of the sequence  $(h_{TX}*h_{RX})(t)|_{\tau T}$. Hence, the SNR of $i$-th sub-carrier is $H[i]^2 \text{ SNR}$. The best modulation scheme for this SNR can be decided from Table \ref{tab:loading_thresholds}.

\begin{figure*}[htbp]
    \centering
    \begin{subfigure}{\textwidth}
        \centering
        \includegraphics[width=\textwidth]{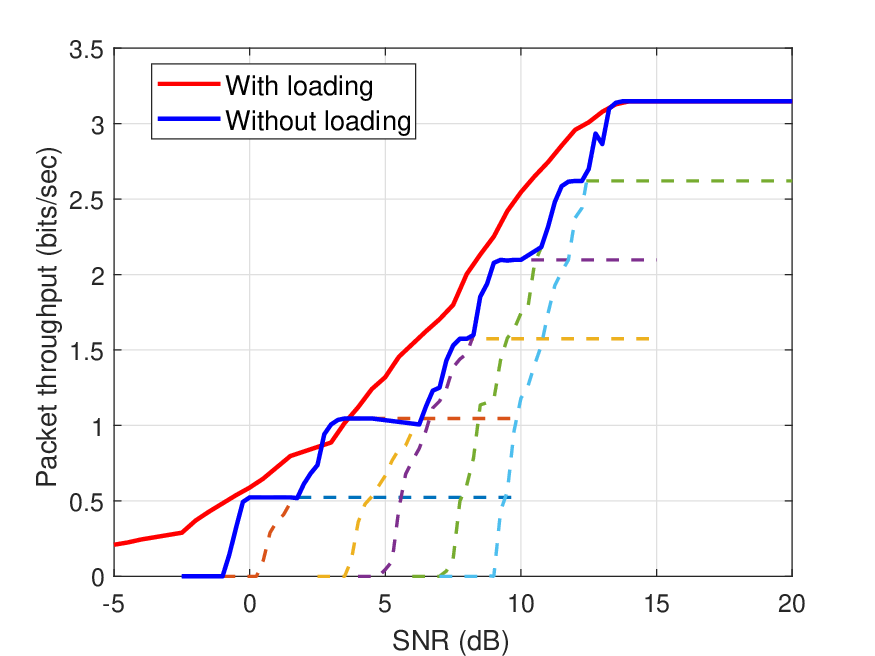}
        \caption{$\tau = 0.9$}
    \end{subfigure}
    \begin{subfigure}{\textwidth}
        \centering
        \includegraphics[width=\textwidth]{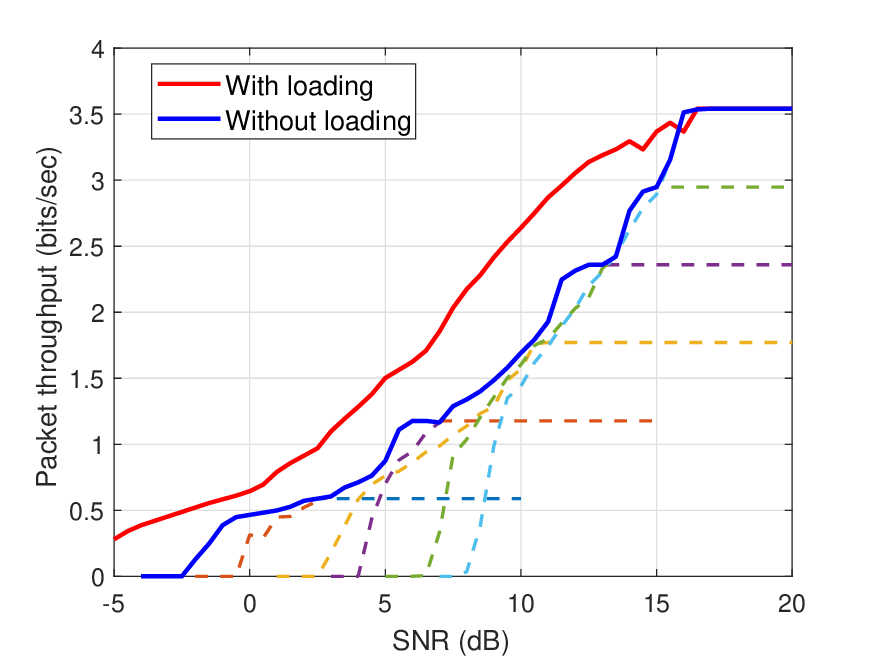}
        \caption{$\tau = 0.8$}
    \end{subfigure}
    \caption{Packet throughput versus SNR for OFDM FTN systems with and without adaptive bit loading.}
    \label{fig:loading}
\end{figure*}

Simulation of OFDM FTN system with  adaptive bit loading was performed. The entries in Table \ref{tab:loading_thresholds} were used to decide the modulation scheme of each sub-carrier. The pseudocode for this simulation is given in Appendix \ref{sec:append_loading}. The resulting packet throughput values are plotted against SNR in Fig. \ref{fig:loading}.
The simulation is done for two values of $\tau=\{0.9, 0.8\}$. For each value of $\tau$, the simulations are performed with and without employing the adaptive loading technique. The throughput obtained with loading is plotted as the solid red curve. The throughput obtained without loading is plotted as the solid blue curve.

The throughput simulation without loading was performed as follows. Firstly, the throughput values corresponding to each modulation scheme were obtained from separate simulations. These values are shown as dotted curves in the figure. The highest throughput values at each SNR was taken among these and the best curve was generated. This best throughput curve is shown as the solid blue curve in the figure.

It can be seen that the throughput values obtained with loading beats the throughput values obtained without loading for almost all values of SNR. The reason for this is the following. In the simulation without loading, same modulation scheme is used for all sub-carriers and this scheme is decided based solely on the average SNR of all sub-carriers. However, there are few sub-carriers that have higher SNR than the average. The transmission rate in these sub-carriers could have been improved by modulating the data on these sub-carriers with a higher rate modulation scheme. 
On the other hand, there are some sub-carriers that have lower SNR than the average. The symbol error rate in these sub-carriers could have been reduced by  modulating the data on these sub-carriers with a lower rate modulation scheme. 
On the contrary, the adaptive loading technique handles each sub-carrier separately. It decides the modulation scheme separately for each sub-carrier and balances the transmission rate and error rate smartly.


\section{Simulation with both Power Allocation and Adaptive Loading}
\label{sec:final_sim}
Simulations employing both power allocation and adaptive loading techniques were performed for OFDM FTN systems. The pseudocode of this simulation is given in Appendix \ref{sec:append_combined}. In this simulation, the power allocated to each sub-carrier was first calculated using the water-filling algorithm. After that, considering this reallocated power, the SNR of each sub-carrier was calculated. Now, from these SNR values, the modulation scheme of each  sub-carrier was decided by referring to Table \ref{tab:loading_thresholds}. Using these modulation schemes, the data on each sub-carrier was modulated to generate the symbols. These symbols were then sent over an AWGN channel with FTN signalling and the overall throughput was calculated at the receiver. 

\begin{figure}[h]
    \centering
    \includegraphics{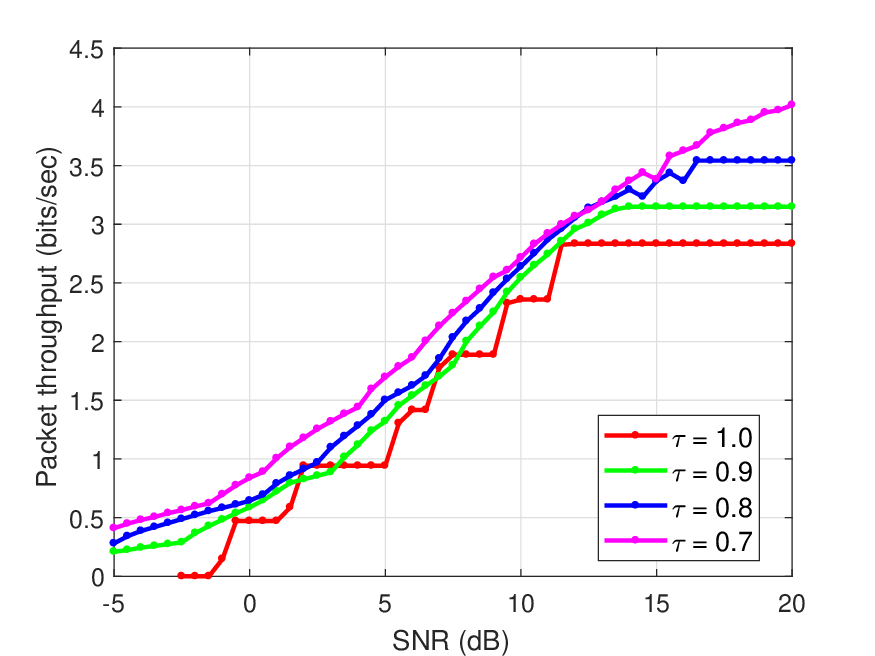}
    \caption{Packet throughput of OFDM FTN system for different values of $\tau$ employing both adaptive loading and power allocation.}
    \label{fig:finalcomp}
\end{figure}

The simulation results plotting the packet throughput values against SNR for different values of $\tau$ is shown in Fig. \ref{fig:finalcomp}.
It can be seen that the packet throughput values are higher for lower values of $\tau$. The reasoning behind this is the following. 

As the value of $\tau$ decreases, symbols get packed closer, resulting the transmission rate to increase. However, as the value of $\tau$ decreases, the ISI also becomes more severe, resulting the symbol error rate to increase. So, there are two opposing effects on the throughput with decreasing value of $\tau$.

The ISI introduced by FTN is handled by the OFDM structure of this system. Further, the combination of power allocation and adaptive loading reduces the severity of ISI by allocating suitable power and modulation schemes to each sub-carriers depending on the local SNR of the sub-carrier. As a result, the effect of higher ISI is less severe than the effect of closer symbol packing. Hence, the overall packet throughput values increase with decreasing values of $\tau$.

This simulation completes the discussion on achieving higher throughput with FTN signalling. This validates the argument made with the use of capacity expressions in Chapter \ref{chap:capacity_perspective} that FTN signalling can provide higher transmission rates than traditional Nyquist signalling.


\chapter{KEY RESULTS AND SUMMARY}
\label{chap:key_results_summary}
In a traditional digital communication system, the rate at which the symbols are spaced is limited by the Nyquist zero-ISI criterion. The method of FTN signalling ignores the zero-ISI criterion and transmits at a faster rate. This introduces ISI in the received signal. Techniques like equalization, precoding, and adaptive loading in OFDM are required to handle this ISI. 

In Chapter \ref{chap:isi_perspective}, the conditions on pulse shape and the time acceleration was derived so that the ISI introduced by FTN signalling is completely invertible. This condition for SRRC modulating pulse was seen to be
\begin{equation}
\label{eq:key_result_limit}
\tau (1+\alpha) > 1
\end{equation}
where, $\tau$ is the time accelration factor of FTN signalling and $\alpha$ is the roll-off factor of the SRRC pulse. When this condition is satisfied, with the help of precoding, detection can be done in a symbol-by-symbol manner without any ISI.

In Chapter \ref{chap:capacity_perspective}, the capacity of FTN system was derived. It was shown that the capacity of FTN system was higher than that of a Nyquist system. The variation of capacity as a function of $\tau$ was discussed. It was shown that, when SRRC modulating pulse is used, the FTN capacity saturates to its highest value and does not increase for values of $\tau$ below the limit in  (\ref{eq:key_result_limit}). This reinforces the condition (\ref{eq:key_result_limit}) derived in Chapter  \ref{chap:isi_perspective}.

In Chapter \ref{chap:waterfilling_loading}, techniques were discussed to increase the throughput of OFDM FTN system. In OFDM FTN systems, the channel gains of each sub-carrier varies according to the modulating pulse shape. The power assigned to each sub-carrier can be varied depending on the corresponding sub-carrier gain. Further, the modulating scheme used in each sub-carrier can also be modified depending on the sub-carrier gain. These techniques were implemented and the throughput simulation results were demonstrated in Chapter \ref{chap:simulation}.


\chapter{FUTURE WORK}
\label{chap:future_work}
The most important challenge in FTN signalling is handling the ISI. As discussed in the thesis, handling the ISI in single-carrier FTN systems is done with the help of precoders and equalizers. These techniques become computationally expensive when severe FTN signalling is used or when the modulation scheme with large constellations are used. In future, computationally efficient equalization algorithms can be explored to equalize the heavy ISI introduced by FTN signalling.

Some of the recent works \citep{kim2018communication, jiang2019turbo, 8052521, Brink:EECS} explore the use of neural networks and deep learning algorithms to design encoders and decoders for communication systems. It has been shown that the neural networks can be trained to closely mimic the trellis structures found in error correction decoders like the turbo decoder and the LDPC decoder. Similar trellis structures are also found in equalization algorithms like the Viterbi algorithm. These results show a potential in designing a computationally efficient deep learning algorithms based on neural networks that can be used to mitigate the heavy ISI introduced by FTN signalling.

In fact, just by using the OFDM modulation, handling the ISI becomes much easier compared to single-carrier systems. By exploring the deep learning techniques along with the OFDM structure, the throughput values can further be improved. Moreover, in an OFDM FTN system, a neural network based encoder and decoder can potentially replace both the power allocation and adaptive loading techniques discussed in this thesis. This can improve the throughput as well as decrease the complexity of the transceiver.

The FTN signalling scheme is a beautiful technical and mathematical problem that has the potential to change the entire picture of the present day communication data rates. By reducing the computational complexity of ISI handling, the FTN signalling can potentially be used in future communication standards to massively improve the data transmission rates. 


\chapter{APPENDIX}
\label{chap:appendix}
\section{PSD of Transmitted Signal}
\label{sec:append_psd}
The PSD of the transmitted signal $\sum_n s[n] h_{TX}(t-nT)$ is given by
\begin{equation}
\text{PSD} = \textbf{E}\left(\left|\frac{1}{\sqrt{T}}\mathcal{F}\left(\sum_n s[n] h_{TX}(t-nT)\right)\right|^2\right)
\end{equation}
where, $\mathbf{E}$ is expectation over the signal ensemble and $\mathcal{F}$ represents the Fourier transform.
\begin{align*}
\text{PSD} &= \frac{1}{T}\textbf{E}\left(\left|\sum_n s[n] \mathcal{F}\left(h_{TX}(t-nT)\right)\right|^2\right) \\
    &= \frac{1}{T}\textbf{E}\left(\left|\sum_n s[n] H(f) e^{-j2\pi fT}\right|^2\right)
\end{align*}
where, $H(f)$ is the Fourier transform of $h_{TX}(t)$. By writing the magnitude in terms of the complex conjugate,
\begin{align*}
\text{PSD} &= \frac{1}{T}\textbf{E}\left(\sum_n s[n] H(f) e^{-j2n\pi fT} \sum_m \overline{s[m]} \overline{H(f)}  e^{j2m\pi fT}\right) 
\end{align*}
where, $\overline{s}$ represents complex conjugate of $s$. Simplifying further,
\begin{align*}
\text{PSD} &=\frac{1}{T} \textbf{E}\left(\sum_n \sum_m s[n]\overline{s[m]} \left|H(f)\right|^2 e^{-j2(n-m)\pi fT}\right)\\
&=\frac{1}{T} \sum_n \sum_m \textbf{E}\left(s[n]\overline{s[m]}\right) \left|H(f)\right|^2 e^{-j2(n-m)\pi fT}.
\end{align*}
Since $s[n]$ are assumed to be i.i.d. and having unit average energy, ergodicity can be employed to write $\textbf{E}\left(s[n]\overline{s[m]}\right) = \delta_{nm}$. Hence,
\begin{align*}
\text{PSD} &= \frac{1}{T}\sum_n \sum_m \delta_{nm} \left|H(f)\right|^2 e^{-j2(n-m)\pi fT} \\
&= \left|H(f)\right|^2/T.
\end{align*}

This shows that the PSD of the transmitted signal is equal to the squared magnitude response $|H(f)|^2/T$ of the modulating pulse $h_{TX}(t)$. Now, if $h_{TX}(t)$ is a sinc pulse, then $H(f)$ is rectangular. Hence, the PSD is also flat.

\section{Comparison of $C_{flat}$ and $C_{non-flat}$}
\label{sec:append_cap_compare}
The capacity expressions of continuous time systems with modulating pulses that are sinc and non-sinc are given by $C_{flat}$ and $C_{non-flat}$ respectively.
\begin{equation}
    C_{flat} = W\log_2 (1+\text{SNR}) \text{ bits/s},
    \label{eq:c_flat_append}
\end{equation}
\begin{equation}
    C_{non-flat} = \int_{-\infty}^{\infty} \log_2\left(1+\text{SNR } W |H(f)|^2\right) df \text{ bits/s}
    \label{eq:c_non_flat_append}
\end{equation}
where, $H(f)$ is the Fourier transform of the non-sinc modulating pulse. Note that non-sinc modulating pulse is assumed to be Nyquist w.r.t $T$ and has bandwidth $W = 1/T$. Since the pulse is square root Nyquist w.r.t. $T$, it satisfies
\begin{equation}
\label{eq:nyq_f}
\frac{1}{T} \sum_{k=-\infty}^{\infty} \left|H\left(f - \frac{k}{T} \right)\right|^2 = 1, \quad \forall f
\end{equation}

Now, in order to show that $C_{flat} < C_{non-flat}$,
\begin{align*}
C_{flat} &= W\log_2 (1+\text{SNR}) \\
&= \int_{0}^{W}\log_2 (1+\text{SNR}) df.
\end{align*}
Using \ref{eq:nyq_f}, 
\begin{align*}
C_{flat} &= \int_{0}^{W}\log_2 (1+\text{SNR}) df \\
&=\int_{0}^{W}\log_2 \left(1+\text{SNR } \frac{1}{T} \sum_{k=-\infty}^{\infty} \left| H\left(f - \frac{k}{T} \right) \right|^2 \right) df \\
&\leq \sum_{k=-\infty}^{\infty} \int_{0}^{W}\log_2 \left(1+\text{SNR } \frac{1}{T} \left| H\left(f - \frac{k}{T}  \right) \right|^2\right) df \\
&=  \sum_{k=-\infty}^{\infty} \int_{-\frac{k}{T}}^{W-\frac{k}{T}}\log_2 \left(1+\text{SNR } \frac{1}{T} \left| H\left(f\right) \right|^2\right) df \\
&=  \sum_{k=-\infty}^{\infty} \int_{-\frac{k}{T}}^{\frac{1}{T}-\frac{k}{T}}\log_2 \left(1+\text{SNR } \frac{1}{T} \left| H\left(f\right) \right|^2\right) df \\
&=  \sum_{k=-\infty}^{\infty} \int_{\frac{k}{T}}^{\frac{k+1}{T}}\log_2 \left(1+\text{SNR } \frac{1}{T} \left| H\left(f\right) \right|^2\right) df \\
&= \int_{-\infty}^{\infty}\log_2 \left(1+\text{SNR } \frac{1}{T} \left| H\left(f\right) \right|^2\right) df \\
&= \int_{-\infty}^{\infty}\log_2 \left(1+\text{SNR } W \left| H\left(f\right) \right|^2\right) df \\
&= C_{non-flat}. 
\end{align*}

This shows that $C_{non-flat} \geq C_{flat}$ in case of non-sinc pulses that satisfy the Nyquist ISI criterion. 

\section{Capacity of FTN System with SRRC Modulating Pulse}
\label{sec:append_cap_ftn_srrc}

The capacity of a FTN system with SRRC modulating pulse is given in (\ref{eq:c_ftn_srrc}) as
\begin{equation}
C_{FTN,SRRC} = \int_{0}^{1/2\tau T}\log_2\left(1+ \text{ SNR } \sum_{k=-\infty}^\infty \frac{1}{T} \mathcal{H}_{RC}\left(f-\frac{k}{\tau T}\right) \right) df \text{ bits/s}.
\end{equation}

In the above expression, the integrand is a function of multiple copies of the CTFT $\mathcal{H}_{RC}$ shifted by $1/\tau T$. Since RC pulse is bandlimited between $-(1+\alpha)/2T$ and $(1+\alpha)/2T$ and both $\tau<1$ and $\alpha < 1$, it can be easily verified that only the first two copies of the $1/\tau T$-shifted CTFT will occur in the integration interval of 0 to $1/2\tau T$. Hence, the expression can be simplified to
\begin{equation}
\label{eq:simple_c_srrrc}
C_{FTN,SRRC} = \int_{0}^{1/2\tau T}\log_2\left(1+ \text{ SNR } \frac{1}{T} \left( \mathcal{H}_{RC}\left(f\right) + \mathcal{H}_{RC}\left(f-\frac{1}{\tau T}\right) \right) \right) df.
\end{equation}

\subsection{For values of $\tau \geq 1/(1+\alpha)$}
To understand the variation of this capacity with the time acceleration factor $\tau$, its derivative is obtained w.r.t. $\tau$ as
\begin{equation*}
\frac{dC_{FTN,SRRC}}{d\tau}=\frac{d}{d\tau} \int_{0}^{1/2\tau T}\log_2\left(1+ \text{ SNR } \frac{1}{T} \left( \mathcal{H}_{RC}\left(f\right) + \mathcal{H}_{RC}\left(f-\frac{1}{\tau T}\right) \right) \right) df.
\end{equation*}
Using the Leibniz integral rule,
\begin{align*}
\frac{dC_{FTN,SRRC}}{d\tau} &= \int_{0}^{1/2\tau T} \frac{d}{d\tau} \log_2\left(1+ \text{ SNR } \frac{1}{T} \left( \mathcal{H}_{RC}\left(f\right) + \mathcal{H}_{RC}\left(f-\frac{1}{\tau T}\right) \right) \right) df \\
 &\hspace{0.5cm}+ \frac{d}{d\tau}\left(\frac{1}{2\tau T}\right) \log_2\left(1+ \text{ SNR } \frac{1}{T} \left( \mathcal{H}_{RC}\left(\frac{1}{2\tau T}\right) + \mathcal{H}_{RC}\left(\frac{1}{2\tau T}-\frac{1}{\tau T}\right) \right) \right),
\end{align*}
\\[-2.5\baselineskip]
\begin{align*}
\hspace{1.3cm} &= \int_{0}^{1/2\tau T} \frac{ \text{ SNR } \mathcal{H}_{RC}'\left(f-\frac{1}{\tau T}\right)\frac{1}{\tau^2 T^2}}{1+ \text{ SNR } \frac{1}{T} \left( \mathcal{H}_{RC}\left(f\right) + \mathcal{H}_{RC}\left(f-\frac{1}{\tau T}\right) \right)} df \\
 &\hspace{0.5cm}-\frac{1}{2\tau^2 T} \log_2\left(1+ \text{ SNR } \frac{1}{T} \left( \mathcal{H}_{RC}\left(\frac{1}{2\tau T}\right) + \mathcal{H}_{RC}\left(\frac{-1}{2\tau T}\right) \right) \right)
\end{align*}
where, $\mathcal{H}_{RC}'(f)$ is the derivative of $\mathcal{H}_{RC}(f)$ w.r.t. $f$. Since $\mathcal{H}_{RC}(f)$ is symmetric, further simplifying,
\begin{multline}
\label{eq:dc_dt}
\frac{dC_{FTN,SRRC}}{d\tau} = \frac{\text{SNR}}{\tau^2 T^2}\int_{-1/\tau T}^{-1/2\tau T} \frac{  \mathcal{H}_{RC}'\left(f\right)}{1+ \text{ SNR } \frac{1}{T} \left( \mathcal{H}_{RC}\left(f + \frac{1}{\tau T}\right) + \mathcal{H}_{RC}\left(f\right) \right)} df \\
 -\frac{1}{2\tau^2 T} \log_2\left(1+ \text{ SNR } \frac{2}{T} \left( \mathcal{H}_{RC}\left(\frac{1}{2\tau T}\right) \right) \right).
\end{multline}

Now, to see the sign of this expression, the following is observed. The minima of the expression $\mathcal{H}_{RC}\left(f\right) + \mathcal{H}_{RC}\left(f + 1/\tau T\right)$ occurs at the frequency at which the two curves $\mathcal{H}_{RC}\left(f\right)$ and $\mathcal{H}_{RC}\left(f + 1/\tau T\right)$ meet, and that frequency is $f=-1/2\tau T$. This result arrives because of the symmetry of the CTFT of the RC pulse. It can also be proved even more robustly by explicitly writing out the expressions for the CTFT of the RC pulse and finding the minima. By using this result,
\begin{align*}
\mathcal{H}_{RC}\left(f\right) + \mathcal{H}_{RC}\left(f + 1/\tau T\right) &\geq \mathcal{H}_{RC}\left(-1/2\tau T\right) + \mathcal{H}_{RC}\left(-1/2\tau T + 1/\tau T\right) \\
&\geq \mathcal{H}_{RC}\left(1/2\tau T\right) + \mathcal{H}_{RC}\left(1/2\tau T\right) \\
&\geq 2\mathcal{H}_{RC}\left(1/2\tau T\right);\quad \forall f.
\end{align*}
\\[-2\baselineskip]
Using the above result, 
\\[-1.75\baselineskip]
\begin{align*}
\mathcal{H}_{RC}\left(f\right) + \mathcal{H}_{RC}\left(f + 1/\tau T\right) &\geq 2\mathcal{H}_{RC}\left(1/2\tau T\right);\quad \forall f, \\
1+ \text{ SNR } \frac{1}{T} \left( \mathcal{H}_{RC}\left(f + \frac{1}{\tau T}\right) + \mathcal{H}_{RC}\left(f\right) \right) &\geq 1+ \text{ SNR } \frac{2}{T} \mathcal{H}_{RC}\left(1/2\tau T\right);\quad \forall f, \\
\frac{1}{1+ \text{ SNR } \frac{1}{T} \left( \mathcal{H}_{RC}\left(f + \frac{1}{\tau T}\right) + \mathcal{H}_{RC}\left(f\right) \right)} &\leq \frac{1}{1+ \text{ SNR } \frac{2}{T} \mathcal{H}_{RC}\left(1/2\tau T\right)} ;\quad \forall f.
\end{align*}
By multiplying both sides with positive $\mathcal{H}_{RC}'(f)$ and integrating both sides, 
\begin{multline*}
\int_{-1/\tau T}^{-1/2\tau T} \frac{  \mathcal{H}_{RC}'\left(f\right)}{1+ \text{ SNR } \frac{1}{T} \left( \mathcal{H}_{RC}\left(f + \frac{1}{\tau T}\right) + \mathcal{H}_{RC}\left(f\right) \right)} df  \\
\leq \int_{-1/\tau T}^{-1/2\tau T} \frac{\mathcal{H}_{RC}'\left(f\right)}{1+ \text{ SNR } \frac{2}{T} \mathcal{H}_{RC}\left(1/2\tau T\right)}  df
\end{multline*}
\begin{equation*}
\hspace{6.15cm}= \frac{\mathcal{H}_{RC}\left(-1/2\tau T\right) - \mathcal{H}_{RC}\left(-1/\tau T\right)}{1+ \text{ SNR } \frac{2}{T} \mathcal{H}_{RC}\left(1/2\tau T\right)}
\end{equation*}
\begin{equation*}
\hspace{5.5cm}= \frac{\mathcal{H}_{RC}\left(1/2\tau T\right) - \mathcal{H}_{RC}\left(1/\tau T\right)}{1+ \text{ SNR } \frac{2}{T} \mathcal{H}_{RC}\left(1/2\tau T\right)}
\end{equation*}
\begin{equation}
\hspace{4.95cm}\leq \frac{\mathcal{H}_{RC}\left(1/2\tau T\right)}{1+ \text{ SNR } \frac{2}{T} \mathcal{H}_{RC}\left(1/2\tau T\right)}.
\label{eq:before_log_ineq}
\end{equation}
Now, the following logarithmic inequality is utilized.
\begin{equation}
\frac{x}{1+x} \leq \log(1+x),\quad \forall x>-1.
\end{equation}
Applying this on the right hand side of the inequality (\ref{eq:before_log_ineq}),
\begin{multline*}
\frac{\text{SNR}}{T}\int_{-1/\tau T}^{-1/2\tau T} \frac{  \mathcal{H}_{RC}'\left(f\right)}{1+ \text{ SNR } \frac{1}{T} \left( \mathcal{H}_{RC}\left(f + \frac{1}{\tau T}\right) + \mathcal{H}_{RC}\left(f\right) \right)} df \\
\leq \frac{1}{2}\frac{\text{SNR}\frac{2}{T}\mathcal{H}_{RC}\left(1/2\tau T\right)}{1+ \text{ SNR } \frac{2}{T} \:\mathcal{H}_{RC}\left(1/2\tau T\right)}
\end{multline*}
\begin{equation*}
\hspace{8.85cm} \leq \frac{1}{2} \log\left(1+\text{SNR}\frac{2}{T}\mathcal{H}_{RC}\left(\frac{1}{2\tau T} \right) \right)
\end{equation*}
or
\begin{multline*}
\frac{\text{SNR}}{\tau^2T^2}\int_{-1/\tau T}^{-1/2\tau T} \frac{  \mathcal{H}_{RC}'\left(f\right)}{1+ \text{ SNR } \frac{1}{T} \left( \mathcal{H}_{RC}\left(f + \frac{1}{\tau T}\right) + \mathcal{H}_{RC}\left(f\right) \right)} df \\
\leq \frac{1}{2\tau^2T} \log\left(1+\text{SNR}\frac{2}{T}\mathcal{H}_{RC}\left(\frac{1}{2\tau T} \right) \right)
\end{multline*}
Now, from (\ref{eq:dc_dt}), it can be concluded that $dC_{FTN,SRRC}/d\tau \leq 0,\; \forall \tau \geq 1/(1+\alpha)$.

\subsection{For values of $\tau < 1/(1+\alpha)$}

Since $\mathcal{H}_{RC}(f)$ is bandlimited between $-(1+\alpha)/2T$ and $(1+\alpha)/2T$, the second copy is bandlimited between  $1/\tau T-(1+\alpha)/2T$ and $1/\tau T + (1+\alpha)/2T$. For $(1+\alpha) < 1/\tau$, the starting frequency of second copy is  $1/\tau T-(1+\alpha)/2T > 1/\tau T-1/2\tau T = 1/2\tau T$. That means the second copy is zero throughout the interval of integration i.e.
\begin{equation}
\mathcal{H}_{RC}\left(f-\frac{1}{\tau T}\right)=0, \quad \text{for } f\in[0,1/2\tau T], \quad\text{when }\tau < 1/(1+\alpha)
\end{equation}

Hence, the capacity expression in (\ref{eq:simple_c_srrrc}) becomes
\begin{equation}
C_{FTN,SRRC} = \int_{0}^{1/2\tau T} \log_2\left(1+ \text{ SNR } \frac{1}{T} \mathcal{H}_{RC}(f)\right) df.
\end{equation}

Using the Leibniz integral rule,
\begin{equation}
\frac{dC_{FTN,SRRC}}{d\tau} = \frac{d}{d\tau}\left(\frac{1}{2\tau T}\right) \log_2\left(1+ \text{ SNR } \frac{1}{T} \mathcal{H}_{RC}\left(\frac{1}{2\tau T}\right) \right) 
\end{equation}
Since $\tau<1/(1+\alpha)$, $1/2\tau T > (1+\alpha)/2\tau$, hence, $\mathcal{H}_{RC}(1/2\tau T) = 0$. Plugging this into the above expression, it can be concluded that  $dC_{FTN,SRRC}/d\tau = 0,\; \forall \tau < 1/(1+\alpha)$.

\section{Pseudocodes of the Simulations}
\label{sec:append_pseudo}
\subsection{Power allocation with water-filling algorithm}
\label{sec:append_waterfilling}
The pseudo code of the water-filling algorithm to get power allocation is shown in Algorithm \ref{algo:wf}.  

\begin{algorithm}
\caption{Water-filling algorithm}\label{algo:wf}
\begin{algorithmic}[1]
\Function{PowerAllocation}{$\text{vector }\textit{channel\_snrs}$}
\State $\textit{channel\_snrs} \gets \text{sort}(\textit{channel\_snrs})$
\State $\textit{N} \gets \text{length of }\textit{channel\_snrs}$
\For{$i = 0 \text{ to } N-1$} 
\State $\textit{temp} \gets 0$
\For{$j = i \text{ to } N-1$}
\State $\textit{temp} \gets \textit{temp} + 1/\textit{channel\_snrs}[j]$
\EndFor
\State $\textit{snr\_threshold} \gets (N-i)/(N + \textit{temp});$
\If{$\textit{channel\_snrs}[i-1] \leq \textit{snr\_threshold} < \textit{channel\_snrs}[i]$}
\State \text{break}
\EndIf
\EndFor
\State $\textit{channel\_powers} \gets 1/\textit{snr\_threshold} - 1/\textit{channel\_snrs}$
\ForAll{$\textit{channel\_powers} < 0$}
\State $\textit{channel\_powers} \gets 0$
\EndFor
\State $\textit{channel\_powers} \gets \text{unsort}(\textit{channel\_powers})$
\State \Return $\textit{channel\_powers}$
\EndFunction
\end{algorithmic}
\end{algorithm}

Here, the input $\textit{channel\_snrs}$ is the vector of SNR values of each sub-carrier and the output $\textit{channel\_powers}$ is the vector of power values allocated to each sub-carrier. Also, $\textit{channel\_snrs}[-1]$ is taken to be 0 and the function unsort in line 14 is the inverse operation of the sort function in line 2.

\subsection{Obtaining baseline throughput values for adaptive loading}
\label{sec:append_baseline}
The pseudo code to obtain the baseline throughput values of Fig. \ref{fig:basethroughput} is shown in Algorithm \ref{algo:baseline}.

\begin{algorithm}
\caption{Obtaining the baseline throughput values of Fig. \ref{fig:basethroughput}}\label{algo:baseline}
\begin{algorithmic}[1]
\Function{GetBaselineThroughput}{$\text{vectors }\textit{SNR\_list},\textit{modulation\_list}$}
\State $\textit{throughput} \gets \text{empty vector}$
\For{$\textbf{each }\textit{SNR}\textbf{ in }\textit{SNR\_list}$}
\For{$\textbf{each }\textit{modulation }\textbf{in }\textit{modulation\_list}$}
\State $b \gets \text{Generate random bits}$
\State $s \gets \textit{modulation}(b)$
\State $n \gets \text{Generate white Gaussian noise}$
\State $n \gets n \times 10^{-\textit{SNR}}$
\State $r \gets s+n$
\State $y \gets \textit{demodulation}(r)$
\State $\textit{throughput} \gets \text{append}(\textit{throughput}, \text{count}(y == b))$
\EndFor
\EndFor
\State \Return $\textit{throughput}$
\EndFunction
\end{algorithmic}
\end{algorithm}

Here, the input $\textit{SNR\_list}$ is the list of SNR values for which the simulation needs to be done and $\textit{modulation\_list}$ is the list of modulation types for which the throughput values are needed. The output $\textit{throughput}$ contains all the resultant throughput values.

\subsection{Adaptive bit loading}
\label{sec:append_loading}
The pseudo code for the implementation of adaptive bit loading is shown in Algorithms \ref{algo:loading} and \ref{algo:loading_demod}. Algorithm \ref{algo:loading} shows the modulation at transmitter with adaptive loading. Algorithm \ref{algo:loading_demod} shows the demodulation at receiver with adaptive loading. 

\begin{algorithm}
\caption{Adaptive bit loading - modulation at the transmitter}\label{algo:loading}
\begin{algorithmic}[1]
\Function{BitLoadingTx}{$\text{vectors }\textit{channel\_snrs}, \textit{bits}$}
\State $N \gets \text{length of }\textit{channel\_snrs}$
\State $\textit{symbols} \gets \text{empty vector}$
\State $i \gets 0$
\While{$i < \text{length}(\textit{bits})$}
\For{$j = 0 \text{ to }N-1$}
\If{$\textit{channel\_snrs}[j] == 0$}
\State $\textit{symbols} \gets \text{append}(\textit{symbols},0)$
\Else
\State $\textit{modulation} \gets \text{modulation type for }\textit{channel\_snrs}[j]\text{ from Table \ref{tab:loading_thresholds}}$
\State $m \gets \text{bits per symbol of }\textit{modulation}$
\State $\textit{symbols} \gets \text{append}(\textit{symbols}, \textit{modulation}(\textit{bits}[i:i+m]))$
\State $i \gets i+m$
\EndIf
\EndFor
\EndWhile
\State \Return $\textit{symbols}$
\EndFunction
\end{algorithmic}
\end{algorithm}

\begin{algorithm}
\caption{Adaptive bit loading - demodulation at the receiver}\label{algo:loading_demod}
\begin{algorithmic}[1]
\Function{BitLoadingRx}{$\text{vectors }\textit{channel\_snrs}, \textit{symbols}$}
\State $N \gets \text{length of }\textit{channel\_snrs}$
\State $\textit{bits} \gets \text{empty vector}$
\State $i \gets 0$
\While{$i < \text{length}(\textit{symbols})$}
\For{$j = 0 \text{ to }N-1$}
\If{$\textit{channel\_snrs}[j] \neq 0$}
\State $\textit{demodulation} \gets \text{demodulation type for }\textit{channel\_snrs}[j]\text{ from Table \ref{tab:loading_thresholds}}$
\State $\textit{bits} \gets \text{append}(\textit{bits}, \textit{demodulation}(\textit{symbols}[i]))$
\EndIf
\State $i \gets i+1$
\EndFor
\EndWhile
\State \Return $\textit{bits}$
\EndFunction
\end{algorithmic}
\end{algorithm}

Here, $\textit{bits}$ is the vector of information bits, $\textit{symbols}$ is the vector of symbols modulated according to the adaptive bit loading method, and $\textit{channel\_snrs}$ is the vector of SNR values of each sub-carrier. This SNR value of each sub-carrier is assumed to be constant throughout. Also, in line 10 of the  Algorithm \ref{algo:loading} and line 8 of the Algorithm \ref{algo:loading_demod}, the modulation type is decided from Table \ref{tab:loading_thresholds} as discussed in Section \ref{sec:loading_sim}.

\subsection{OFDM FTN simulation with both power allocation and adaptive bit loading}
\label{sec:append_combined}
The pseudo code to obtain the throughput values with both power allocation and adaptive bit loading technique is shown in Algorithm \ref{algo:final}.

\begin{algorithm}
\caption{OFDM FTN simulation with power allocation and bit loading}\label{algo:final}
\begin{algorithmic}[1]
\Function{FtnOfdmSimulation}{$\text{vector }\textit{SNR\_list},\textit{OFDM\_length}, \textit{pulse},T,\tau$}
\State $N \gets \textit{OFDM\_length}$
\State $h \gets \textit{pulse}\text{ sampled at }\tau T$
\State $H \gets N\text{-point DFT of }h$
\State $\textit{throughput} \gets \text{empty vector}$
\For{$\textbf{each }\textit{SNR}\textbf{ in }\textit{SNR\_list}$}

\State $\textit{channel\_snrs} \gets \textit{SNR} \times H^2$
\State $P \gets \textsc{PowerAllocation}(\textit{channel\_snrs})$
\State $\textit{channel\_snrs} \gets P\times\textit{channel\_snrs}$

\State $b \gets \text{Generate random bits}$

\State $s \gets \textsc{BitLoadingTx}(\textit{channel\_snrs}, b)$
\State $s \gets \sqrt{P}\times s$
\State $s \gets \text{IFFT and CP addition of }s$

\State $n \gets \text{Generate white Gaussian noise}$
\State $n \gets n \times 10^{-\textit{SNR}}$

\State $r \gets \text{conv}(h,s)+n$
\State $r \gets \text{CP removal and FFT of }r$
\State $r \gets r/\sqrt{P}H$

\State $y \gets \textsc{BitLoadingRx}(\textit{channel\_snrs}, r)$
\State $\textit{throughput} \gets \text{append}(\textit{throughput}, \text{count}(y == b))$
\EndFor
\State \Return $\textit{throughput}$
\EndFunction
\end{algorithmic}
\end{algorithm}

Here, $\textit{SNR\_list}$ is the list of SNR values for which the simulation needs to be done,  $\textit{OFDM\_length}$ is the number of bits per OFDM symbol, $\textit{pulse}$ is the modulating pulse, $T$ is the rate w.r.t. to which the modulating pulse is Nyquist, and $\tau$ is the time acceleration factor of FTN signalling. Here, wherever multiplication with a vector is performed, it is taken to be elementwise multiplication. Also, wherever operations like IFFT, FFT, or element-wise multiplication is performed on a vector, the vector is reshaped accordingly.


\begin{singlespace}
\bibliography{refs}
\end{singlespace}


\listofpapers

\begin{enumerate}  
\item \textbf{A. Jain, S. Chadaga, N. Seshadri, and R. D. Koilpillai}  \newblock
 Faster-Than-Nyquist Signalling with Constraints on Pulse Shapes
  \newblock {\em 2019 IEEE 30th Annual International Symposium on Personal, Indoor and Mobile Radio Communications (PIMRC)},
  1-6, (2019).
\end{enumerate}  

\end{document}